\documentclass{aa}  
\usepackage[flushleft]{threeparttable}
\usepackage{graphicx}
\usepackage{txfonts}
\usepackage{xcolor}
\usepackage{hyperref}
\usepackage{soul}
\begin{document}

\def\simgt{\lower.5ex\hbox{$\; \buildrel > \over \sim \;$}}
\def\simlt{\lower.5ex\hbox{$\; \buildrel < \over \sim \;$}}
\def\etal{{\it et al.}}
\def\msun{M$_\odot$}
\def\Msun{M$_\odot$}
\def\rsun{R$_\odot$}
\def\Teff{$T_{\rm eff}~$}
\def\teff{$T_{\rm eff}$}
\def\BMV{$B-V$~}
\def\EBV{$E_{B-V}$}
\def\dv{$\Delta V$~}
\def\Mc{$M_\rm c$~}
\def\X{$X$~}
\def\FeH{${\rm [Fe/H]}$~}
\def\feh{${\rm [Fe/H]}$}
\def\Y{$Y$~}
\def\Z{$Z$~}
\def\Lto{$L_{\rm TO}$~}
\def\Hp{$H_{\rm p}$~}
\def\logL{$\log (L/L_{\odot})$~}
\def\Mv{$M_{\rm v}$~}
\def\M{$M$}
\def\R{$R$}
\def\alfamlt{$\alpha_{\rm MLT}$~}
\def\alfaatm{$\alpha_{\rm atm}$~}
\def\alfa2d{MLT--$\,\alpha^{\rm 2D}$~}
\def\afa{$\alpha$~}
\def\taufot{$\tau_{\rm ph}$~}
\def\logg{$\log (g)$}
\def\Logg{$\log (g)$~}
\def\Dnu{$\Delta\nu$~}
\def\dnu{$\Delta\nu$~}
\def\Deltanu{$\Delta\nu$~}
\def\deltanu{$\Delta\nu$}
\def\numax{$\nu_{\rm max}$}
\def\Numax{$\nu_{\rm max}$~}
\def\numaxm{\nu_{\rm max}~}
\def\Dnum{\Delta_{\rm nu}~}
\def\vmic{v$_{\rm mic}$~}
\def\kepler{\mbox{\textit{Kepler}}~}
\def\Space{\mbox{SP\_Ace}~}

\title{Masses and ages for metal-poor stars}

\subtitle{A pilot program combining asteroseismology and high-resolution spectroscopic follow-up of RAVE halo stars \footnote{Based on data collected during ESO program 099.D-0913(a)}}
  
\author{M.~Valentini
          \inst{1}
          \and C.~Chiappini\inst{1} \and D.~Bossini\inst{2,3} \and A.~Miglio\inst{4,5}  \and G.~R.~Davies\inst{4,5} \and B.~Mosser\inst{6}  \and Y.~P.~Elsworth\inst{4,5} \and S.~Mathur\inst{7,8}  \and Rafael A. Garc\'{i}a\inst{9,10} \and L.~Girardi\inst{2} \and T.~S.~Rodrigues\inst{2} \and M.~Steinmetz\inst{1} \and A. Vallenari\inst{2}}         

   \institute{Leibniz-Institut f\"ur Astrophysik Potsdam (AIP), An der Sternwarte 16, 14482 Potsdam, Germany
\and
Osservatorio Astronomico di Padova, INAF, Vicolo dell'Osservatorio 5, I-35122 Padova, Italy 
\and
Instituto de Astrof\'isica e Ci$\hat{e}$ncias do Espa\c{c}o, Universidade do Porto, CAUP, Rua das Estrelas, 4150-762 Porto,Portugal
\and
School of Physics and Astronomy, University of Birmingham, Edgbaston, Birmingham, B15 2TT, UK
\and
Stellar Astrophysics Centre, Department of Physics and Astronomy, Aarhus University,  DK-8000 Aarhus C, Denmark 
\and
LESIA, Observatoire de Paris, PSL Research University, CNRS, Universit\'{e} Pierre et Marie Curie, Universit\'{e} Paris Diderot, 92195 Meudon, France
\and
Departamento de Astrof\'{i}sica, Universidad de La Laguna, E-38206 Tenerife, Spain
\and
Instituto de Astrof\'{i}sica de Canarias, C/ V\'{i}a L\'{a}ctea s/n, La Laguna, E-38205 Tenerife, Spain
\and
IRFU, CEA, Universit\'{e} Paris-Saclay, F-91191 Gif-sur-Yvette, France
\and
AIM, CEA, CNRS, Universit\'{e} Paris-Saclay, Universit\'{e} Paris Diderot, Sorbonne Paris Cit\'{e}, F-91191 Gif-sur-Yvette, France
}

   \date{Received ??? ??, ????; accepted ??? ??, ????}
  
  \abstract{Very metal-poor halo stars are the best candidates for being among the oldest objects in our Galaxy. Samples of halo stars with age determination and detailed chemical composition measurements provide key information for constraining the nature of the first stellar generations and the nucleosynthesis in the metal-poor regime.}  
{Age estimates are very uncertain and are available for only a small number of metal-poor stars. Here we present the first results of a pilot program aimed at deriving precise masses, ages and chemical abundances for metal-poor halo giants using asteroseismology, and high-resolution spectroscopy.}
{We obtained high-resolution UVES spectra for four metal-poor RAVE stars observed by the K2 satellite. Seismic data obtained from K2 light curves helped improving spectroscopic temperatures, metallicities and individual chemical abundances. Mass and ages were derived using the code PARAM, investigating the effects of different assumptions (e.g. mass loss, [$\alpha$/Fe]-enhancement). Orbits were computed using Gaia DR2 data.}
{The stars are found to be {\it normal} metal-poor halo stars (i.e. non C-enhanced), with an abundance pattern typical of old stars (i.e. $\alpha$ and Eu-enhanced), and with masses in the 0.80-1.0 \Msun~range. The inferred model-dependent stellar ages are found to range from 7.4 to 13.0 Gyr, with uncertainties of $\sim$ 30\%-35\%. We also provide revised masses and ages for metal-poor stars with {\it Kepler} seismic data from APOGEE survey and a set of M4 stars.}
{The present work shows that the combination of asteroseismology and high-resolution spectroscopy provides precise ages in the metal-poor regime. Most of the stars analysed in the present work (covering the metallicity range of  [Fe/H] $\sim$ $-$0.8 to $-$2 dex), are very old $>$9 Gyr (14 out of 19 stars ), and all of them are older than $>$ 5 Gyr (within the 68 percentile confidence level).}

   \keywords{Stars - fundamental parameters -- Asteroseismology -- Stars - abundances}

   \maketitle
%
\section{Introduction}

The Milky Way halo is a key component to understand the assembly history of our Galaxy. The halo is composed by stars that were accreted during mergers as well as stars that formed in-situ \citep[e.g.,][]{Helmi1999,Helmi2018}, and is suggested to be one of the oldest component of our Galaxy,
(e.g., \citealp{Jofre2011, Kalirai2012, Kilic2019}). In addition, metal-poor halo giant stars enshrine information on when star formation began, on the nature of the first stellar generation and on the chemical enrichment time-scale in the Galactic halo (\citealp{Cayrel2001,Chiappini2013,Frebel2015}). A comprehensive understanding of the Galactic halo can be obtained only when combining precise stellar chemical abundances, kinematics, and ages. While detailed chemical information can be obtained via high-resolution spectroscopic analysis {and precise kinematics is being provided by astrometric missions like Gaia}, the determination of reliable stellar ages (i.e. ages that are precise and unbiased), is still a challenging task, especially in the case of red giants. 

Before the confirmation of solar-like oscillations in red-giant stars \citep{DeRidder2009}, ages had been estimated only for a limited sample of nearby field stars, either by model-dependent techniques such as isochrone fitting, or empirical methods such as nucleo-cosmo-chronometry. The age determination via the classic isochrone-fitting method has always been hampered by the fact that in the red-giant locus the isochrones clump together, which leads to a large degeneracy. This degeneracy leads to age uncertainties easily above 80\% for the oldest stars (e.g., \citealp{daSilva2006, Feuillet2016}). The few metal-poor field halo stars with a better age determination than the isochrone fitting uses the nucleo-cosmo-chronometry technique (mostly derived using the Th-232 and U-238 ratio), and these indicate old ages (\citealp{Cayrel2001,Cowan2002,Hill2002,Sneden2003,Ivans2006,Frebel2007,Hill2017, Placco2017}). These old ages seem to confirm the expectations that metal-poor halo objects are among the oldest objects in our Galaxy. Although the nucleo-cosmo-chronometry method is more precise than isochrone fitting in the case of red giants, it is not a viable solution for all stars. The method requires high-resolution and high signal-to-noise (SNR) spectra in the blue region of the spectrum (SNR$>$300 at $\sim$390 nm), and high r-process enhancement in order to allow for the presence of strong, and sufficiently measurable, U and Th lines. 

Asteroseismology of red giant stars has, in recent years, demonstrated to provide precise masses for such stars, and therefore ages 
 (\citealp{Casagrande2016, Anders2016, SilvaAguirre2018}). Solar-like oscillations are commonly summarised by two parameters: \Deltanu
 (average frequency separation) and \Numax (frequency of maximum oscillation power). These two quantities provide precise mass (precision of about
  10\%) and radius (precision of about 3\%), using the so-called seismic scaling relations, and an additional information on stellar temperature
 (\teff) (\citealp{Miglio2013,Casagrande2014,Pinsonneault2014}). Since for red giants the stellar masses are a good proxy for stellar age, it is possible to determine a  model-dependent age with a precision that can be better than 30\% depending on the quality of the seismic 
 information \citep{Davies2016AN}. More precise ages, error $\sim$15\%, can be obtained via Bayesian methods combining seismic information with 
 Gaia data and information on the stellar evolutionary stage \citep[][and references therein]{Rodrigues2017}.

Since the age determination using asteroseismology relies on the mass-age relation that red giants follow, this means that the method is biased by any event that changes the stellar mass, as, for example, mass accretion from a companion or stellar mergers \citep[blue stragglers, or stars rejuvenated by accretion, e.g. ][]{Boffin2015} or mass-loss.  One way to look for mass accretion events from a companion is to look for radial velocity, photometric variations, or chemical signs of accretion (e.g. high carbon and s-process enhancements - \citealp{Beers2005, Abate2015}). The effect of mass-loss can be minimised by looking at stars in the low-RGB phase, where the effect of mass loss are smaller compared to red-clump stars \citep{Anders2016}.
   \begin{figure}
   \centering
   \includegraphics[width=1\columnwidth]{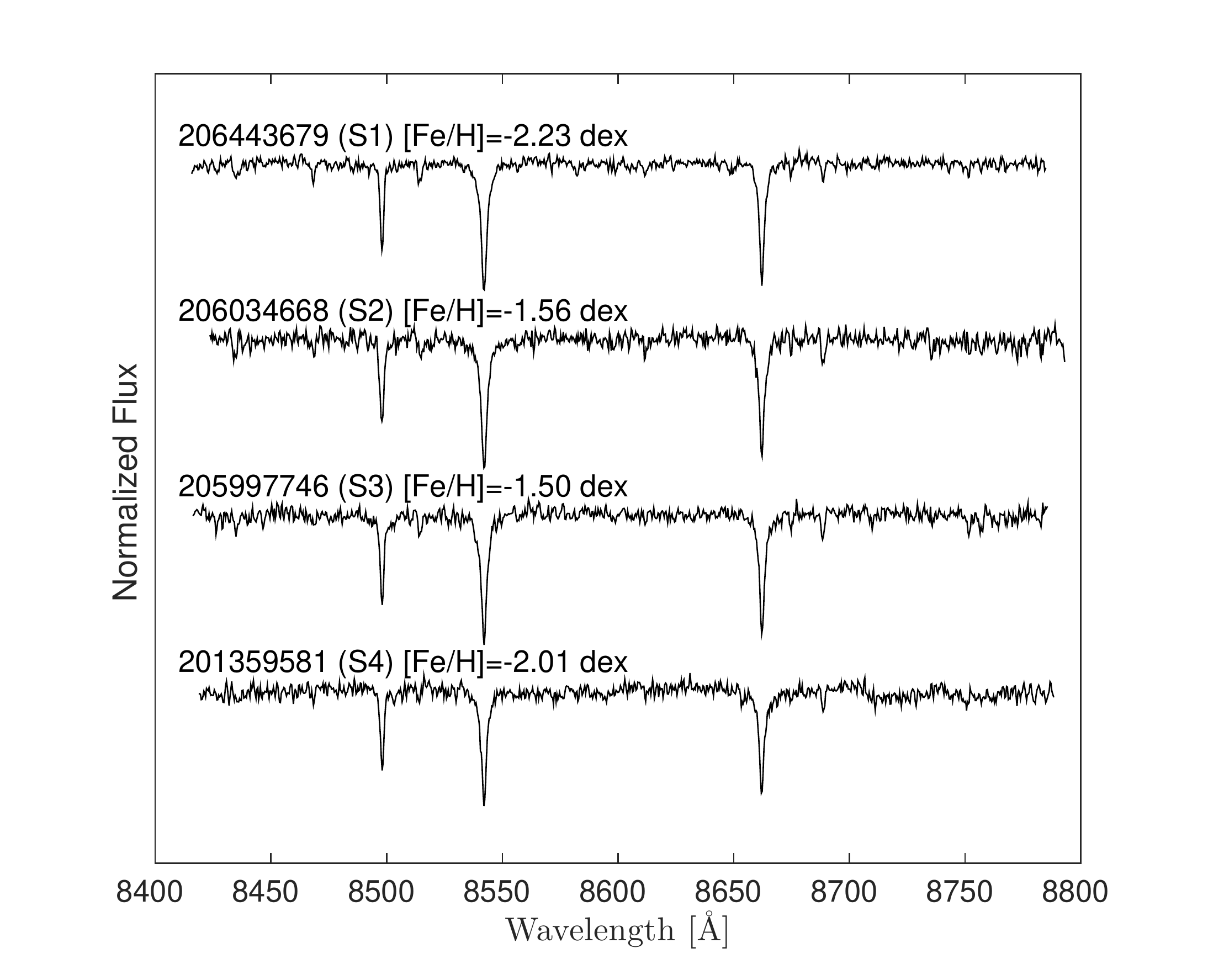}
   \caption{RAVE spectra of the 4 metal-poor stars presented in this paper. Spectra are normalised and corrected for radial velocity, the Fe content labeled comes from the analysis of RAVE spectra using the same method as in \cite{Valentini2017}.}
              \label{Fig:spectra}
    \end{figure}
The first study to determine masses for a sample of metal-poor halo giants with both seismic information (from {\it Kepler},  \citealp{Borucki2010}) and chemistry from high-resolution APOGEE \citep{Majewski2015} spectra, was the one of \cite{Epstein2014}. The authors used scaling relations at face value and reported masses larger (M$>$~1~\Msun) than what would be expected for a typical old population. Similar results were obtained by \cite{Casey2018}, also using scaling relations for three metal-poor stars. These findings led to the need for further tests of the use of asteroseismology in the low metallicity regime. \citet{Miglio2016}, analysed a group of red giants in the globular cluster M4 (\FeH $= -$1.10 dex and [$\alpha$/Fe]=0.4 dex) with seismic data from K2 mission \citep{Howell2014}, and found low seismic masses compatible with the old age of the cluster, hence suggesting that seismic masses and radii estimates would be reliable in the metal-poor regime provided a correction to the \Deltanu scaling relation is taken into account for red giant branch (hereafter RGB) stars. The correction presented in \citet{Miglio2016} is a correction theoretically motivated, based on the computation of radial mode frequencies of stellar modes. 

In this work, we present a first set of four stars, identified as metal poor ([Fe/H] $\sim -2$ dex) in the RAVE survey, for which we have seismic information from the K2 mission and high-resolution spectra.
The paper is organised as follows: in Sec.~\ref{Sect:data} we describe how the stars have been selected and observed. The seismic light curve analysis and the determination of atmospheric parameters and abundances from stellar spectra are described in Sec.~\ref{Sect:analysis}. In Sec.~\ref{Sect:massage} we derive radii, masses and ages for our stars using both PARAM and scaling relations. We recompute masses for the \citet{Epstein2014} and M4 \citep{Miglio2016} samples. We also analyse the offsets and uncertainties introduced by different seismic pipelines, erroneous assumptions in temperature, [$\alpha$/Fe]-enhancements, and mass loss. Distances and orbits of the stars are derived in Sec~\ref{Sect:DandO}, using Gaia-DR2 parallaxes and proper-motions. In Sec.~\ref{Sec:discussionindiv} we discuss each of the four RAVE stars in light of their chemistry, age and orbital properties. Sec.~\ref{Sect:conclusions} summarises our conclusions and provide an outlook.
 
   \begin{figure}
   \centering
   \includegraphics[width=1\columnwidth]{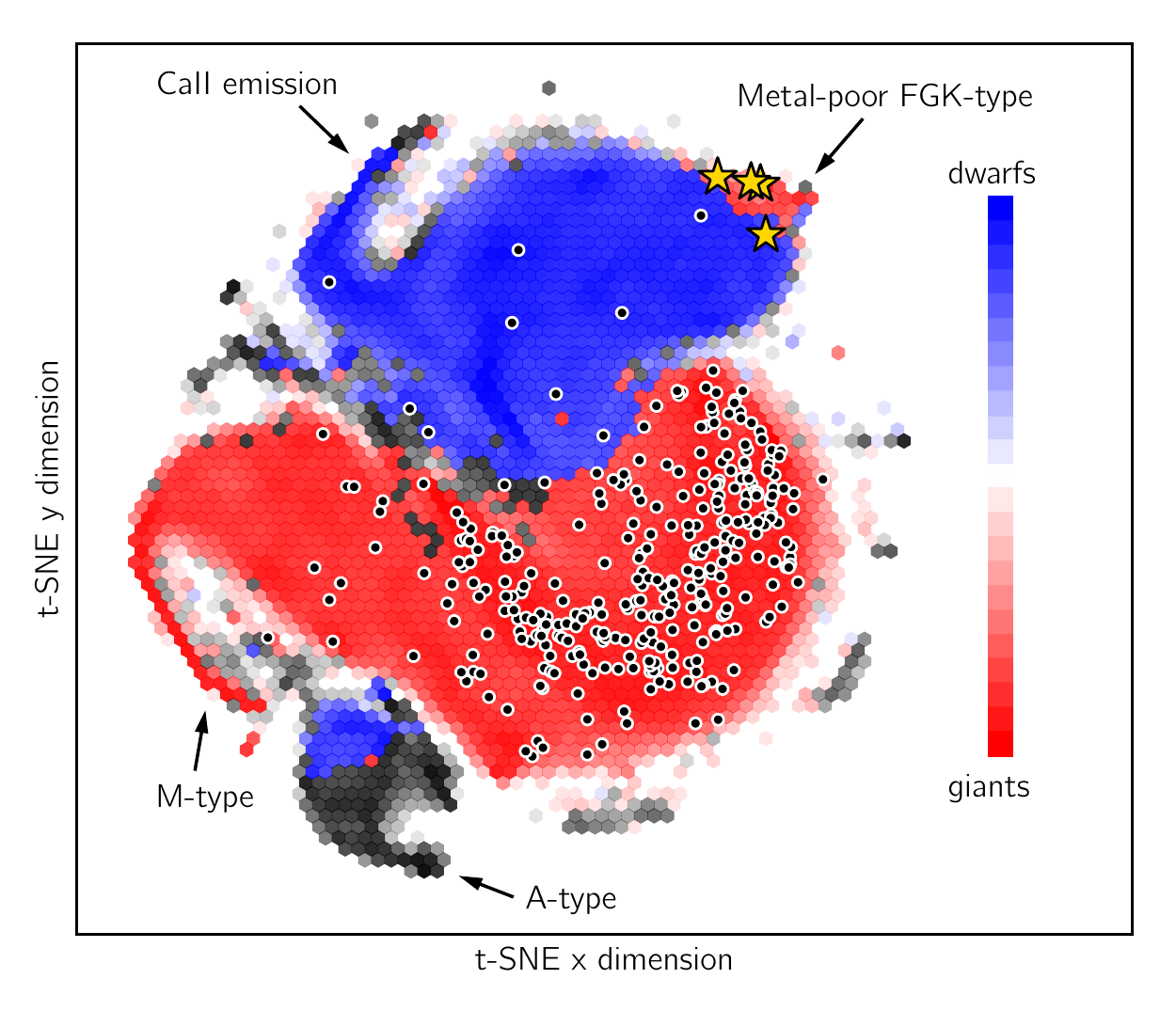}
   \caption{t-SNE projection of $\sim$420,000 RAVE spectra. The scaling in both direction is arbitrary,
therefore the units on the axes are omitted. The colour scale corresponds to the gravity of the
stars as computed by \citet{Kunder2017}. Giants are shown in red and dwarfs in blue. Lighter shaded hexagons include fewer stars than darker ones. Over-plotted black dots indicate locations of RAVE stars in K2 Campaigns 1,3. Illustrated as stars are the RAVE-K2 objects studied in the present work, which fall in the metal-poor locus of the diagram.}
              \label{Fig:Tsne}%
    \end{figure}

\section{Observations}
\label{Sect:data}

\subsection{K2}
Targets analysed in this works belong to K2 mission campaigns 1 and 3. The K2 Campaign 1 field (C1), centred at RA 11 h 35 m 46 s DEC $+$01$^\circ$ 25' 00'' (l=265, b=$+$58), was observed from 30 May 2014 to 21 August 2014, and contains one metal-poor RAVE star. The K2 Campaign 3 field (C3), centred at RA 22 h 26 m 40 s DEC $-$11$^\circ$ 25' 02'' (l=51, b=$-$52), was observed from 14 November 2014 to 03 February 2015, contains three RAVE metal-poor giants. RAVE targets were observed as part of the \textquotedblleft The K2 Galactic Archaeology Program Campaign\textquotedblright (C1-C3 proposal GO1059, and described in \citealp{Stello2015}). 

Light curves were obtained using the same approach as described in Section~3 of \citet{Valentini2017}. 

\subsection{Target selection}

In C1 and C3 K2 fields there are a total of 376 RAVE targets for which solar-like oscillations have been detected. Following the joint spectroscopic and seismic analysis described in \citet{Valentini2017} we identified four stars expected to have metallicities \FeH$\leq$ $-$1.5 dex. The spectra of the metal-poor targets are visible in Fig.\ref{Fig:spectra}.

RAVE spectra cover a narrow spectral interval (8410-8795 \AA) at intermediate resolution (R$\sim$7,500), that combined with the low  metallicity of the targets (few detectable lines, as  visible in Fig.~\ref{Fig:spectra}) make the traditional spectroscopic analysis challenging: the atmospheric parameters may suffer of degeneracies and offsets. Using the t-SNE projection (\citet{MAt2017}, we confirmed that the four stars were, indeed, metal poor. The t-SNE projection \citep{van2008visualizing} is an algorithm that, when applied to spectra, provides a low-dimensional projection of the spectrum space and isolates objects that present similar morphology. In our case, as visible in Fig.~\ref{Fig:Tsne}, metal-poor stars clump in the upper-left region of the projection. In the figure $\sim$420,000 RAVE spectra with SNR $>$ 10 are projected, with the RAVE stars in K2 C1 and C3 represented as empty circles. The four stars that fall into the very metal-poor island (top right) are the metal poor giants analysed in the present work.

\begin{figure*}
\centering
\includegraphics[width=2.\columnwidth]{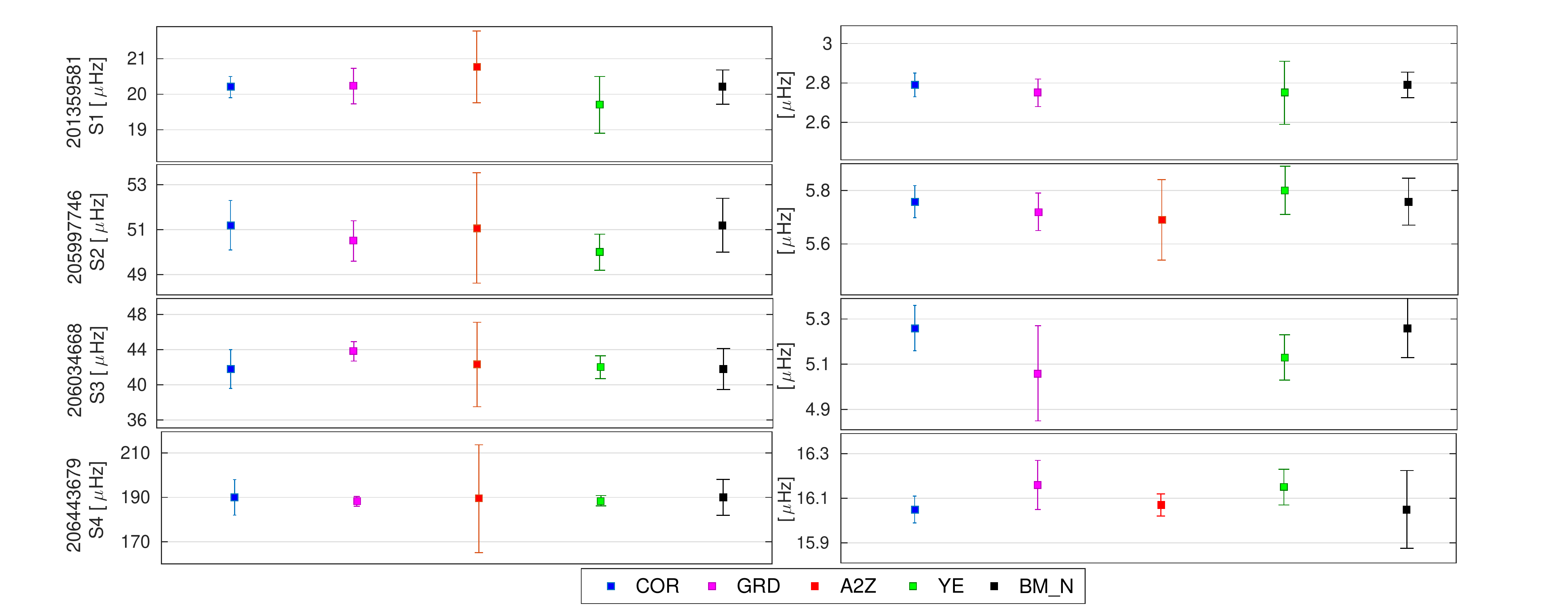}
\caption{\Dnu and \numax as measured by different pipelines. Each colour (blue, magenta, red and green) corresponds to a different pipeline (COR, GRD, A2Z and YE - see Appendix). The values plotted in black correspond to a further test using COR with inflated uncertainties (BM$\_$N). }
\label{Fig:seismopipelines}%
\end{figure*}
    
\subsection{Gaia DR2}
The four stars are in Gaia DR2 (\citealp{Gaia2016,Gaia2018}). Parallaxes, proper motions and flags are listed in Table~\ref{Tab:data}. The \texttt{duplicated\_source} flag is listed as Dup.
\begin{table*}
\caption{Gaia DR2 data and seismic data for the 4 RAVE metal-poor stars studied here. In this work we adopted \Dnu and \Numax from COR pipeline and investigated the effect of adopting errors computed considering the dispersion among four different seismic pipelines (COR, GRD, YE, A2Z), here identified as BM\_N seismic values.}
\label{Tab:data}
\centering          
\begin{tabular}{lcccc}     
\hline\hline  
               &        S1   &      S2            &       S3          &   S4                \\ \hline

GAIA DR2 data  &                    &                    &                   &                     \\ \hline
Gaia source ID &3602288924850161792 &2596851370212990720 &600175713555136256 & 2622975976942392320 \\
$\varpi$       [mas] & 0.4621 & 0.4764  & 0.6027 & 1.3793  \\
$\varpi$ error [mas]& 0.0880 & 0.0543 & 0.0386 & 0.0434  \\
pmra           [mas/yr]& $-$51.3896  & 16.7891  &   $-$24.2069 &   32.4331 \\
pmra error     [mas/yr]&  0.1695  &  0.4889  &  0.0632  &  0.0791  \\
pmdec          [mas/yr]&  $-$4.5454  &  $-$4.5019  & $-$48.9613  &  0.8501 \\
pmdec error    [mas/yr]&  0.0719 &  0.1795  &  0.0592  &  0.0713  \\
Dup.           &  1 & 0 & 0 & 0 \\
Astrom. exc.   &   0.1474; 8.895 &  0; 0      &  0; 0      &  0; 0      \\
Priam flag     &  100001&  100002&  100001&  100001\\ \hline
K2 data& &    &  &  \\ \hline
EpicID         &201359581           &205997746           &206034668          &206443679            \\ 
Kp                   &    10.96    &   12.46    &   11.65    &   12.15    \\
Campaign             & C1 &  C3 & C3   & C3 \\ 
\dnu$_{\rm COR}$    [$\mu$Hz] &  2.79$\pm$0.06  &  5.76$\pm$0.06  &  5.26$\pm$0.10   &  16.05$\pm$0.06    \\
\numax$_{\rm COR}$  [$\mu$Hz] &  20.20$\pm$0.30 &  51.20$\pm$1.10  &  41.80$\pm$2.20  &  190.00$\pm$8.00  \\
\dnu$_{\rm BM\_N}$  [$\mu$Hz] &   2.79$\pm$0.65  & 5.76$\pm$0.88  &  5.26$\pm$0.12  &  16.05$\pm$0.17  \\
\numax$_{\rm BM\_N}$ [$\mu$Hz] &  20.20$\pm$0.48  &  51.20$\pm$1.20  &  41.80$\pm$2.34  & 190.00$\pm$8.03  \\ \hline
\end{tabular}
\end{table*}


Star S1 (Epic ID: 201359581) has a \texttt{duplicated$\_$source} flag = true, meaning that this source presented more than one detection and only one entry was kept. This means that the star had observational or processing problems, leading to possible erroneous astrometric or photometric solution. This same star has an \texttt{astrometric$\_$excess$\_$sigma} $\geq$ 2 that, combined with \texttt{astrometric$\_$excess$\_$noise} flag $>$ 0, indicates large astrometric errors and an untrustworthy solution. For this same star the Gaia DR2 radial velocity has an error of 5.17~km/s, hence larger than the $\sim$~0.8 km/s expected for a star of that temperature and brightness. 

Star S2 (Epic ID: 205997746) has a \texttt{Priam$\_$flag} indicating a {\it silver} photometry quality and a lower quality in the temperature, radius and luminosity solutions (while the rest of the stars in the sample have a better, {\it golden}, photometry quality).

For S1 (201359581), we did not consider the ages and masses derived by taking into account the Gaia DR2 information. In addition, we consider the solutions for S2 (205997746) of lower quality respect to the other 2 stars, S3 (206034668) and S4 (206443679). We will use the Gaia DR2 proper motions when computing orbits for our stars in Section~\ref{Sect:DandO}, with the exception of S1, for which we will use UCAC-5 \citep{UCAC5} proper motions.

Gaia DR2 parallax, $\varpi$, can be used for deriving the surface gravity:
\begin{equation}
\label{Eq:parlogg}
\begin{split}
\log (g) _{\varpi}= \log (g)_\odot + 4 \log \left(\frac{T_{\rm eff}}{{\rm T}_{{\rm eff},\odot}}\right)+\log \left( \dfrac{m}{m_{\odot}}\right) +\\+0.4 \left( m_{V}+5 -5 \log (1/\varpi)-3.2 (E(B-V)) + BC - M_{{\rm bol},\odot} \right)
\end{split}
\end{equation}
We derived \logg$_\varpi$ for the stars of our sample, assuming the bolometric correction (BC) as in \citet{BC2018} and \citet{BC2014}, using Ks magnitudes and assuming stellar masses of 0.9 \msun~ and spectroscopic (UVES) temperatures. Errors were calculated via propagation of uncertainties and varying stellar masses from 0.8 to 2.2 \Msun\ (a typical red giant star mass range). We also took into account the effect of the different offsets in the $\varpi$, considering the zero point correction \citep{Lindegren2018} and the offset pointed out by \citet{Zinn2018}: thus we considered an offset effect that varies $\varpi$ within ($\varpi -$0.3) and ($\varpi +$0.2). Resulting gravities and their uncertainties are listed together with the stellar parameters obtained from spectroscopy (see next Sections) in Table~\ref{Tab:hratmpar}.

\subsection{High-resolution spectra}

UVES high resolution spectra of our targets were collected in the period 99D, using UVES-CD 3 set-up \citep{UVES2000}, program ID: 099.D-0913(A). Spectra have a resolving power of $\sim$110,000 and cover $\sim$4170-6200\AA spectral range. Observing date, exposure time and SNR of spectra are listed in Table~\ref{Tab:ESO}. 

\begin{table*}
\caption{Coordinates and set-up of the ESO-UVES observations of the stars. The SNR listed is the one calculated in the all spectral range.}
\label{Tab:ESO}
\centering          
\begin{tabular}{llcccccc}     
\hline\hline       
ID &  RA & DEC  & JD middle & Set-up & Exp. time & SNR \\ 
   &   [deg]  &   [deg]   &           &        &    [s]       &      \\ \hline
S1 & 178.650541 &  $-$1.56250& 57863.10762427578 & CD3 & 1200 & 60\\
S2 & 339.990916 & $-$14.88894& 57941.21053530200 & CD3 & 3000 & 78\\
S3 & 333.817541 & $-$13.83519& 57889.39524668500 & CD3 & 1300 & 70\\
S4 & 338.755333 &  $-$5.90969& 57950.41013379906 & CD3 & 2600 &100\\
\hline
\end{tabular}
\end{table*}
\section{Data analysis}
\label{Sect:analysis}


\subsection{Seismic Data}
Very metal poor stars typically have large radial velocities that induce a Doppler shift of observed frequencies. Although small, this shift can be larger than the precision on asteroseismic frequencies. In this work we use the average seismic parameters \Dnu and \numax. Because \Dnu is a frequency difference and because the precision on \Numax is much lower than for individual mode frequencies the Doppler correction does not need to be applied to asteroseismic average parameters \citep{Davies2014}.
 
In order to quantify the impact of the different seismic inputs on the estimates of the mass and age of our stars, we first considered the \deltanu~ and  \numax~ measurements coming from four different seismic pipelines:
\begin{itemize}
\item {\bf COR}: It is the method adopted for CoRoT and \kepler stars (\citealp{Mosser2009, Mosser2011}). In a first step, the
average frequency separation $\Delta\nu$, is measured from the autocorrelation of the time series computed as the Fourier spectrum
of the filtered Fourier spectrum of the signal. The significance of the result is checked using a statistical test based on the H0 hypothesis. 
\item {\bf GRD}: This pipeline is based on fitting a background model to the data \citep{Davies2016}. The model is a  model H \citep{Kallinger2014}, comprised of two Harvey profiles, a Gaussian oscillation envelope, and an instrumental noise background. For the estimate of \Numax the central frequency of the Gaussian component is considered. The median and the standard deviations are used to summarise the normal-like posterior probability density for \numax. To estimate the average frequency separation a model was fitted to the power spectrum \citep{Davies2016AN}.
\item {\bf YE}: This is a three stages approach. First, a signal-to-noise ratio spectrum (SNR) in function of frequency is created by dividing the power spectrum by a heavily smoothed version of the raw power spectrum. The second step consists in using a combination of H0 and H1 hypothesis for detecting oscillation power in segments of the SNR spectrum. If a segment shows detection of oscillations power, then \Numax and \Dnu are detected as a third step (\citealp{Hekker2010, Elsworth2017}).
\item {\bf A2Z}: A first estimate of \Dnu was done using the same method as COR. \Numax is measured by fitting a Gaussian on top of the background to the power spectrum. Then \Dnu is recomputed from the power spectrum of the power spectrum and by considering only the central orders of the spectrum centred on the highest radial mode (\citealp{Mathur2010, Mathur2011}). Differently from the previous pipelines, this one measured a value for \Dnu only for 2 of the 4 targets and provided significantly larger error bars for \numax.
\end{itemize}

We then checked that the different pipelines were in agreement for the four stars, as showed in Fig.~\ref{Fig:seismopipelines}. 
As we are dealing with a small number of stars and since the four pipelines are in agreement, we can perform a star-by-star analysis of the goodness of the seismic values. From a visual inspection, as visible in  Appendix~A, it appears that: {\it i)} A2Z pipeline is providing 
\Dnu with very large uncertainties; {\it ii)} YE and GRD pipelines provide a \numax value that appears shifted respect to the expected value, 
for star S1 and S2 respectively (see Appenfix Fig.~\ref{Fig:numax}). As shown by previous works (e.g. \citealp{Lillo2014, Perez2016}), using 
individual frequencies for deriving \Dnu is more precise than the method presented above. The individual frequencies fitting exercise 
is difficult to perform for K2 light-curves, because of the short duration of the K2 runs. For this reason the use of the universal pattern is preferred, as in \citet{Mosser2011}, which uses the detailed information of the whole oscillation pattern \citep{Mosser2011}. This dedicated analysis provides refined values of the global seismic parameters, with smaller uncertainties. This choice is justified also by the tests performed in \citet{Hekker2012}. For these reasons we have therefore adopted \Dnu from the COR pipeline as our preferred value. 

An additional test has been performed, for RGB stars in the $\alpha$-rich  APOGEE-{\it Kepler} \citep[APOKASC][]{Pinsonneault2018} sample:  individual mode frequencies has been measured for $\approx$1,000 stars and then a comparison between \Dnu measured from individual frequencies with the \Dnu measured by COR pipeline had been performed. A small ($\lesssim$ 1\%) difference between \Dnu as determined by COR, and \Dnu determined from individual radial-mode frequencies is found (Davies et al., in preparation), supporting our choice for COR values. This is also relevant because the \Dnu determined from individual mode frequencies is closer to the \Dnu given in the stellar models adopted in PARAM, the tool used in this work for deriving mass, radii, and ages. We additionally considered the seismic values from GRD pipeline, which has error bars in \Dnu and \Numax compatible with the COR pipeline and with the data quality (see more details in Appendix~\ref{SeismoChoice}).

For having a better comprehension of the impact of the use of a {\it global error} coming from considering all the pipelines we also adopted a fifth set of \Dnu and \numax (BM\_N), where the \Dnu and \Numax are from the COR pipeline but with inflated errors that consider the dispersion of the pipelines respect to COR values: 
\begin{equation}
\label{Eq:errors}
\sigma^2_{x,{\rm BM\_N}}=\sigma^2_{x,{\rm COR}}+\frac{\sum_{\rm i=GRD,YE,A2Z}(x_i - x_{\rm COR})^2} {3}
\end{equation}
where $x$=\Dnu or \numax. The adopted seismic values, COR and BM\_N, are listed in Table~\ref{Tab:data} (the complete set of seismic values are in Appendix Table~\ref{Tab:allseismo}) and a comparison of the different sets of \Dnu and \Numax is shown Fig.~\ref{Fig:seismopipelines}. 

   \begin{figure}
   \centering
   \includegraphics[width=0.99\columnwidth]{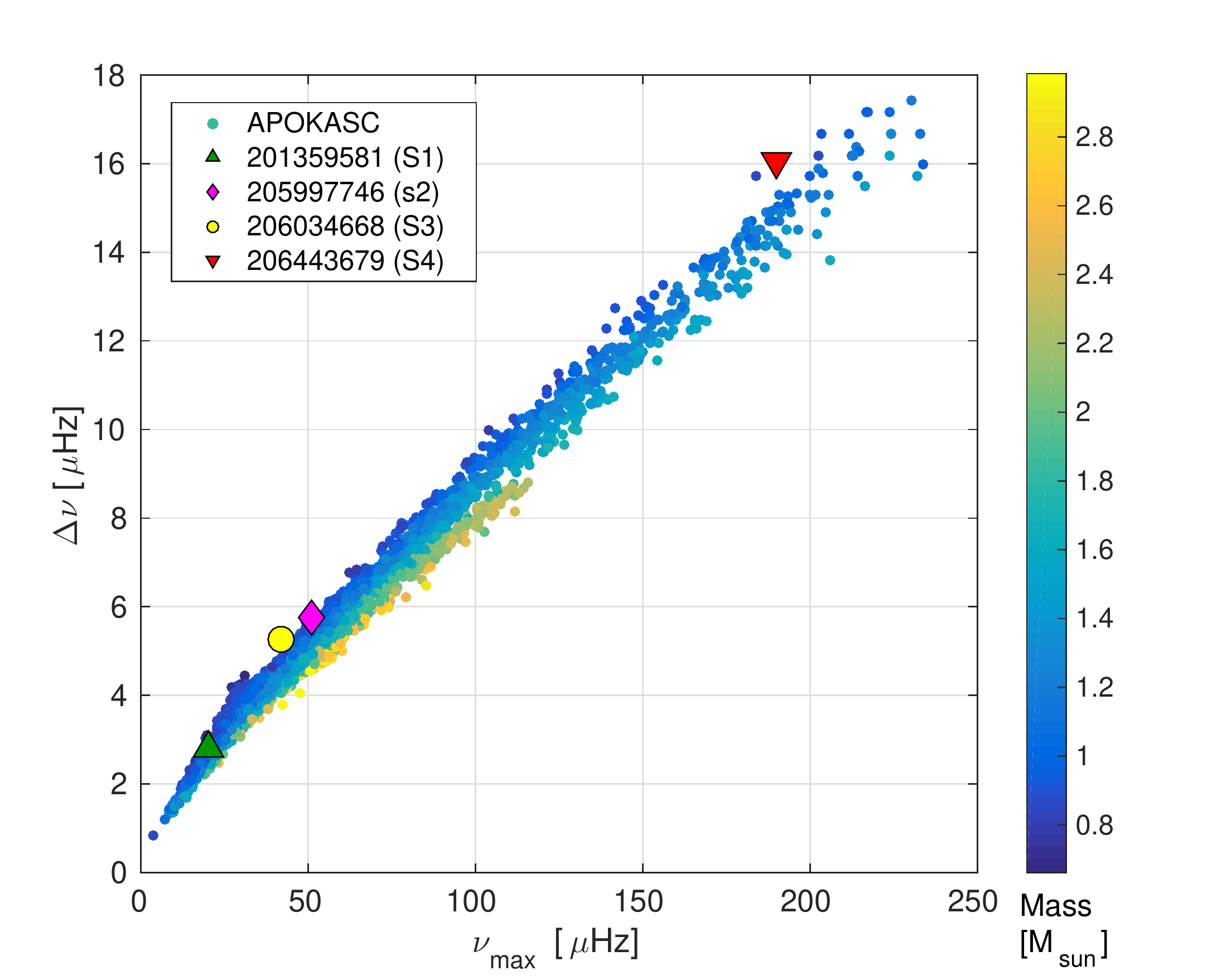}
   \caption{\Dnu and \numax distribution of the 4 stars studied in this work. On the background the \dnu-\numax distribution distribution of the APOKASC sample, colour coded following the mass.}
              \label{Fig:dataseismo}%
    \end{figure}

We compared the \Deltanu and \Numax of our sample with the \Deltanu and \Numax distribution of the APOKASC sample. The high quality of the APOKASC sample makes it the perfect benchmark to provide a first glance on the masses expected for our objects. Fig.~\ref{Fig:dataseismo} shows that our four stars fall in the region where the less massive stars are located.

\subsection{RAVE spectra analysis}

The analysis  of the RAVE spectra has been performed following the method described in \citet[][Sect.~4]{Valentini2017}. We iteratively derived atmospheric parameters by fixing the gravity to the seismic value, \logg$_{\rm S}$. As a starting point for deriving \teff, we used the Infra-Red Flux Method (IRFM) temperature published in RAVE-DR5 \citep{Kunder2017}, allowing for variations as large as 250~K. This analysis was performed using the GAUFRE  pipeline \citep{Valentini2013}. 

The seismic gravity we used is defined as:
\begin{equation}
\label{Eq:sismologg}
\log (g)_{\rm S}= \log (g)_\odot + \log \left(\dfrac{\nu_{\rm max}}{\nu_{{\rm max}, \, \odot}}\right)+ \frac{1}{2} \log \left(\frac{T_{\rm eff}}{{\rm T}_{{\rm eff},\odot}}\right)
\end{equation} 
with the adoption of the following solar values: \numax$_{,\odot}$= 3090 $\mu$Hz, \dnu$_\odot$= 135.1 $\mu$Hz, \logg$_\odot$=4.44 dex, and \teff$_{,\odot}$= 5777 K \citep{Huber2011}.

\begin{table*}
\caption{Radial velocity, atmospheric parameters and abundances of the metal-poor RAVE stars in K2 Campaigns 1 and 3, as derived from RAVE spectra. Temperature and abundances have been derived by fixing the gravity to the seismic value (following the method described in \citet{Valentini2017}) and using RAVE spectra. Abundances were determined under LTE assumptions.}
\label{Tab:stars}
\centering          
\begin{tabular}{llcccccccc}     
\hline\hline  
ID  &Vrad  & \Teff & \Logg & [Fe/H] & [M/H] & [$\alpha$/Fe] & [Mg/Fe] & [Si/Fe] & [Ti/Fe]   \\    
    &[km/s]       & [K] & [dex] & [dex] & [dex] & [dex] & [dex] & [dex] & [dex]   \\  \hline
S1& 77.14$\pm$0.85 &5230$\pm$ 62  & 2.24$\pm$ 0.008 & $-$2.01$\pm$ 0.10 & $-$1.92$\pm$ 0.10 & 0.31$\pm$0.13 & 0.54$\pm$0.15 &1.02$\pm$0.17 & 1.29$\pm$0.18 \\
S2& $-$205.08$\pm$0.87 &5012$\pm$ 81  & 2.58$\pm$ 0.008 & $-$1.50$\pm$ 0.09 & $-$1.29$\pm$ 0.12 & 0.34$\pm$0.16 & -- & 0.54$\pm$0.15 & --   \\
S3& 55.75$\pm$0.69 &4990$\pm$ 93  & 2.57$\pm$ 0.005 & $-$1.56$\pm$ 0.10 & $-$1.24$\pm$ 0.12 & 0.34$\pm$0.15 & 0.76$\pm$0.15 & 0.69$\pm$0.15 & 1.11$\pm$0.15    \\
S4& $-$40.67$\pm$1.45 &5241$\pm$ 90  & 3.17$\pm$ 0.008 & $-$2.23$\pm$ 0.12 & $-$2.23$\pm$ 0.17& 0.23$\pm$0.18 & -- & -- & $-$0.05$\pm$0.17\\
\hline
\end{tabular}
\end{table*}

Atmospheric parameters and abundances derived from RAVE spectra are listed in Table~\ref{Tab:stars}. Abundances were derived under Local Thermodynamic Equilibrium (LTE). The chemical abundances obtained from RAVE spectra suggest that the four stars are $\alpha$-enhanced with [$\alpha$/Fe]$\sim$0.3 dex. On the other hand the individual abundance ratios of [Mg/Fe], [Si/Fe] and [Ti/Fe] are significantly discrepant for the different stars. Notice that the [Fe/H] values reported in Table~\ref{Tab:stars} are not corrected for non-local thermodynamic equilibrium (NLTE) effects. We will return to this point when discussing the abundance ratios obtained from high-resolution spectra.  

\subsection{UVES spectra analysis}
\begin{table*}
\caption{Atmospheric parameters and radial velocities of the stars as obtained from Gaia-DR2 and from high-resolution UVES spectra. The latter values were obtained in two ways: using the classical analysis with MOOG and FeI-FeII equivalent widths (Cl.) or in an iterative way fixing the gravity to the seismic value  (\logg$_{\rm S}$).}
\label{Tab:hratmpar}
\centering          
\begin{tabular}{ll|ccc|ccc|ccc|ccc}     
\hline  \hline 
      &  &   \multicolumn{3}{|c|}{201359581 - S1} &  \multicolumn{3}{|c|}{205997746 - S2} &  \multicolumn{3}{|c|}{206034668 - S3} &  \multicolumn{3}{|c}{206443679 - S4}  \\ \hline
Gaia DR2 &    & &       & $\sigma$   &   &      & $\sigma$&  &     &  $\sigma$ & & &$\sigma$  \\ 
\Teff  & [K]  & &4987   &  $^{+45} _{-87}$     & & 4984 & $^{+30} _{-25}$  & & 5038 & $^{+37} _{-99}$ & & 5121& $^{+125} _{-124}$ \\
v$_{\rm rad}$ & [km/s]& & 70.00   & 5.17 & & $-$204.75 &0.79 & &54.44 & 1.35& & $-$41.23& 2.43 \\
\logg$_\varpi$   & [dex] & & 1.87    & 0.26 & & 2.51      & 0.25 & & 2.39 & 0.25 & & 3.17 & 0.25 \\       \hline
 \multicolumn{2}{c|}{Classical}    & &       & $\sigma$   &  & & $\sigma$  &  & &  $\sigma$  & & & $\sigma$ \\ 
\Teff$_{,\rm Cl}$  & [K]  & &4936   & 63& & 4987 & 78  & & 4890 & 85 & & 5120 & 64 \\ 
\logg$_{\rm Cl}$  & [dex]& &1.98   & 0.20 & & 2.25 & 0.19& & 2.22 & 0.21& & 2.95& 0.20 \\ \hline
 \multicolumn{2}{c|}{With Seismo}    & &       & $\sigma$   &  & & $\sigma$  &  & &  $\sigma$  & & & $\sigma$ \\ 
\Teff  & [K]  & &4850   & 43& & 5020 & 35  & & 4995 & 25 & & 5245& 35 \\ 
\logg$_{\rm S}$  & [dex]& &2.17   & 0.03 & & 2.58 & 0.02& & 2.58 & 0.04& & 3.17& 0.05 \\
vmic   & [km/s]& &2.1   & 0.5  & & 1.80  & 0.5 & & 2.40& 0.4& & 1.8 & 0.5  \\ 
v$_{\rm rad}$   & [km/s]& & 74.63   & 0.11  & & $-$204.80  & 0.08 & & 55.71& 0.08& & $-$41.13 & 0.09  \\ \hline
\end{tabular}
\end{table*}


\begin{table*}
\centering
\caption{Summary of the abundances of the stars of this work. The solar composition adopted is listed in the last column, from \citet{Asplund2009}. Values are corrected for NLTE effects and in case of multiple ions (e.g. FeI and FeII), the mean has been considered.}
\label{Tab:XFe}
\begin{tabular}{lccccccccc}
\hline \hline 
  & \multicolumn{2}{c}{201359581 - S1} & \multicolumn{2}{c}{205997746 - S2} & \multicolumn{2}{c}{206034668 - S3} &  \multicolumn{2}{c}{206443679 - S4} & Sun \\
  {[Fe/H]$_{NLTE}$ }&$-$1.89 $\pm$0.10 &  &  $-$1.33 $\pm$ 0.09 & &  $-$1.42 $\pm$ 0.10 &  &$-$1.94 $\pm$ 0.10 &   &    \\
  \hline
       & [X/Fe] & $\sigma$ & [X/Fe] & $\sigma$ & [X/Fe] & $\sigma$ & [X/Fe] & $\sigma$ & log$\epsilon_{\odot} (\X)$ \\ \hline
 {[C/Fe]} & 0.30   & 0.15 &  -0.18   & 0.10&   0.18   & 0.11 &  0.01  &  0.09 &   8.43 \\
 {[Na/Fe]}& 0.28   & 0.06 &   1.14   & 0.12&   0.25   & 0.05 &  0.41  &  0.08 &   6.24 \\
 {[Mg/Fe]}& 0.45   & 0.11 &   0.63   & 0.05&   0.27   & 0.15 &  0.72  &  0.10 &   7.60 \\
 {[Si/Fe]}& 0.75   & 0.07 &   0.61   & 0.04&   0.62   & 0.10 &  0.81  &  0.10 &   7.51 \\
 {[Ca/Fe]}& 0.48   & 0.05 &   0.42   & 0.10&   0.24   & 0.13 &  0.57  &  0.13 &   6.34 \\
 {[Sc/Fe]}& 0.32   & 0.11 &   0.00   & 0.14&   0.12   & 0.11 &  0.34  &  0.14 &   3.15 \\
 {[Ti/Fe]}& 0.26   & 0.08 &   0.27   & 0.15&   0.17   & 0.10 &  0.59  &  0.10 &   4.95 \\
 {[Cr/Fe]}& 0.01   & 0.22 &   0.05   & 0.11&  $-$0.13 & 0.09 &  0.19  &  0.09 &   5.64 \\
 {[Mn/Fe]}& 0.07   & 0.08 &   0.20   & 0.10&  $-$0.08 & 0.09 &  0.01  &  0.08 &   5.43 \\
 {[Ni/Fe]}& 0.32   & 0.10 &   0.17   & 0.09&  $-$0.03 & 0.11 &  0.19  &  0.07 &   6.22 \\
 {[Cu/Fe]}&$-$0.25 & 0.07 &   0.12   & 0.10&  $-$0.23 & 0.08 &$-$0.26 &  0.10 &   4.25 \\
 {[Zn/Fe]}& 0.30   & 0.11 &   0.26   & 0.11&   0.39   & 0.14 &  0.41  &  0.11 &   4.56 \\
 {[Sr/Fe]}& 0.10   & 0.08 &  $-$0.09 & 0.09&  0.03    & 0.11 &  0.69  &  0.11 &   2.87 \\
 {[Ba/Fe]}& 0.50   & 0.08 &   0.31   & 0.09&   0.92   & 0.10 &  0.83  &  0.13 &   2.18 \\
 {[Eu/Fe]}& 0.80   & 0.07 &   0.41   & 0.08&  0.03    & 0.08 &  0.79  &  0.08 &   0.52 \\
 {[Gd/Fe]}& 0.05   & 0.07 &  $-$0.34 & 0.08&   0.43   & 0.10 &  --    &  --   &   1.07 \\ \hline
\end{tabular}
\end{table*}

   \begin{figure*}[t]
   \centering
   \includegraphics[width=2\columnwidth]{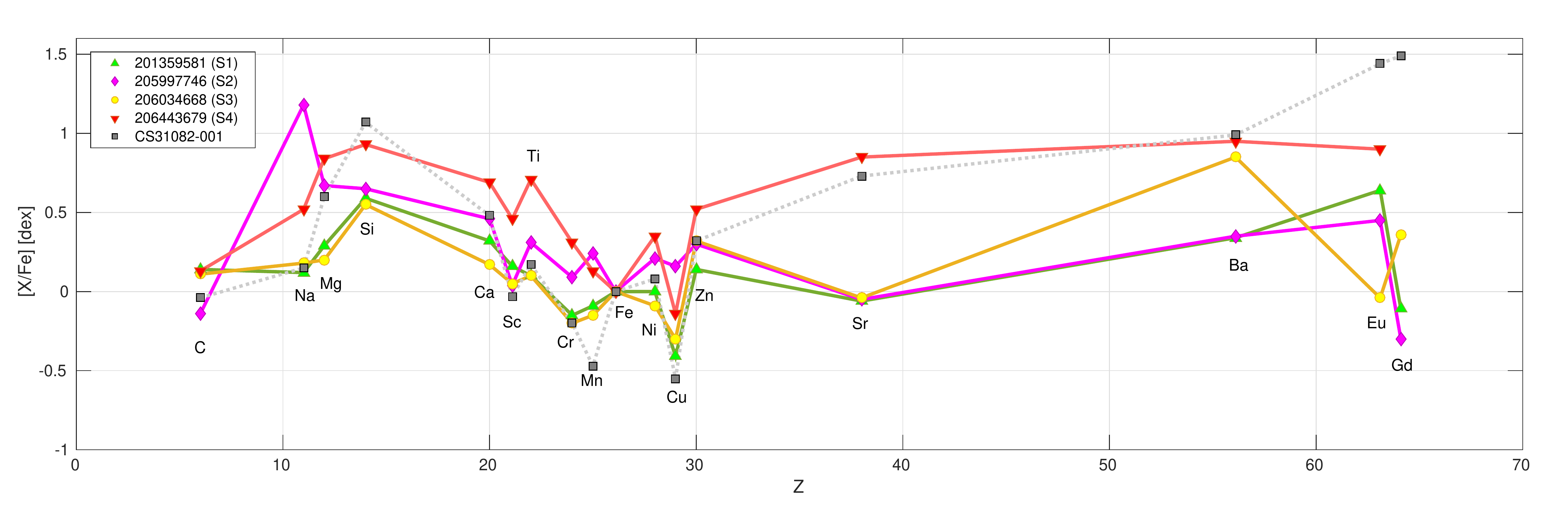}
   \caption{Chemical abundance pattern [X/Fe] for the four metal-poor stars studied in this work. As a reference, the abundances pattern of the r-process enriched star CS 31082-001 are plotted as a dark grey line (abundances from \citet{Roederer2014b}).}
              \label{Fig:abdpattern}%
    \end{figure*}
We analysed the high-resolution UVES spectra using the GAUFRE pipeline for retrieving \teff , \logg , and \FeH iteratively using the seismic information on \logg , using Eq.~\ref{Eq:sismologg}. The analysis was performed with the GAUFRE module GAUFRE$\_$EW, that derives atmospheric parameters via ionisation and excitation equilibrium using the equivalent widths (EW) of FeI and FeII lines, MARCS model atmospheres \citep{MARCS} and the silent version of MOOG 2017\footnote{\url{http://www.as.utexas.edu/$\sim$chris/moog.html}}. For sake of comparison we derived atmospheric parameters also using the classical method (imposing excitation and ionisation equilibrium using FeI and FeII lines), results are listed as \teff$_{,\rm Cl}$ and \logg$_{\rm Cl}$ in Table~\ref{Tab:hratmpar}.  

The error in \Teff was calculated considering the range of \Teff within the Fe~I abundances were independent from the line excitation potential (slope equal to zero) and by varying \Logg and \vmic within errors. The error in \Logg was calculated via propagation of uncertainty when the adopted \Logg was derived using asteroseismology (Eq.~\ref{Eq:sismologg}). When \Logg was measured via the classic method (ionization equilibrium of Fe~I and Fe~II), the uncertainty was derived by varying \Teff, \vmic, and \FeH by their uncertainty, since the values are interdependent. 

Abundances of different chemical elements were derived using MOOG 2017, in the updated version  properly treating Rayleigh scattering \citep{Sobeck2011} \footnote{Code available at: \url{https://github.com/alexji/moog17scat}}. For the abundances analysis an ad-hoc model atmosphere with the same atmospheric parameters found by GAUFRE, was created via interpolation using MARCS models. The linelist was constructed using the linelists in \citet{Roederer2014}, \citet{Hill2002}, implemented, when necessary, with line parameters retrieved from VALD DR4 database (\citealp{VALD1, VALD2, VALD3, VALD4, VALD5}).  The C abundance was derived via fitting the A-X CH band-head at $\sim$4000-4300 \AA. Line parameters were taken from \citet{Masseron2014}. We measured the abundances of the following alpha-elements: Mg, Si, Ca, and Ti. NLTE corrections for Ti are taken from the work of \cite{NLTETi}. In addition we measured the abundances of several iron peak elements (Cr, Mn, Fe, Ni, Cu, Zn, and Ga). For Fe we adopted the line-by-line NLTE corrections provided by \cite{NLTEFe}. NLTE corrections for Mn are taken from \cite{NLTEMn}. Line-by-line corrections for Fe and Mn are taken from a user-friendly interface available online \footnote{Available at the website \url{http://nlte.mpia.de/}}.  As indicator of r-process enrichment we measured abundances of Eu and Gd. As s-process markers we measured Sr and Ba.

Final abundances are listed Table~\ref{Tab:XFe} (for more details see Appendix~\ref{UVESappendix}). The uncertainties on abundances provided in Table~\ref{Tab:XFe} (and in ~\ref{Tab:abundances}) were calculated considering: the internal error of the fit, the errors on \Teff and \logg, and the error on continuum normalisation. The error on the fit is provided by MOOG itself. We computed the impact of \Teff and \logg\ uncertainties by creating different model atmospheres by varying atmospheric parameters within the errors. Error on continuum normalisation has been taken into account by creating, for each stellar spectrum, ten different continuum normalisations and then analysing them. The error listed in Table~\ref{Tab:XFe} is the sum in quadrature of these three different errors. In Figure~\ref{Fig:abdpattern} we compare the abundance pattern of the four RAVE stars with that of CS 31082-001 (dotted grey curve) which is considered to be a typical pure r-process enriched star \citealp{Spite2018}. The  abundances for CS 31082-001 were taken from \citealp{Roederer2014b}. 

   \begin{figure*}
   \centering
   \includegraphics[width=1.8\columnwidth]{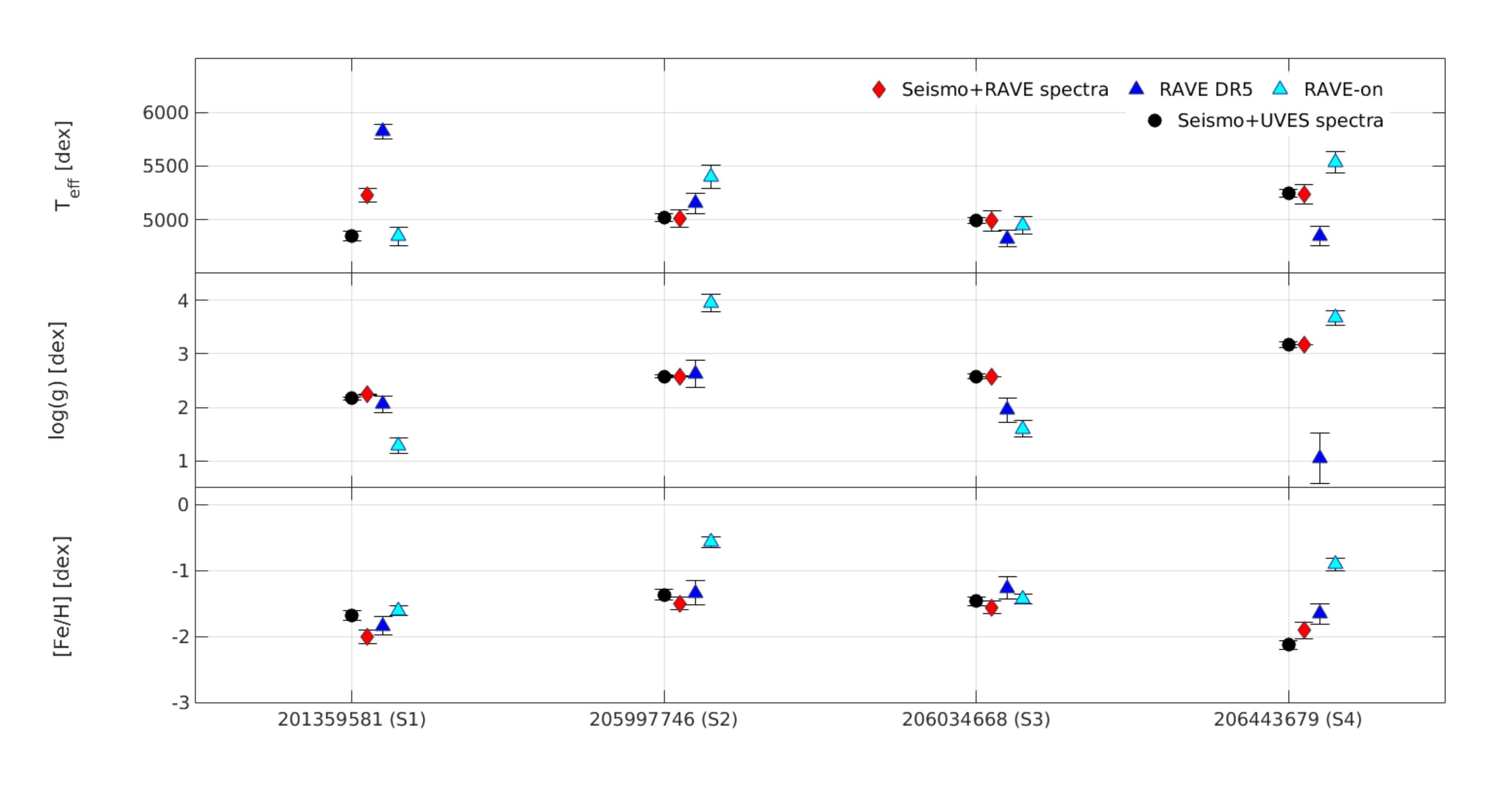}
   \caption{Atmospheric parameters of the sample of metal-poor stars, as taken from literature and this work: RAVE spectra and seismic parameters (red squares), RAVE-DR5 (blue triangles), RAVE-on (cyan triangles) and ESO high-resolution spectra and seismic parameters (black circles). }
              \label{Fig:datalit}
    \end{figure*}

The four stars are clearly enhanced in core collapse (SN type II) nucleosynthetic products (such as Mg, Si, and Eu), as one would expected to be the case for old stars. However, the range in $\alpha$-enhancement is very large, and it is not correlated with metallicity. S1, S2 and S4 can be classified as r-I stars (i.e. stars with 0.3 $\leq$ [Eu/Fe] $\leq$ 1 and [Ba/Eu]$<$, \citealp{Christlieb2004}), while S3 is clearly Ba-enhanced. The low C-enhancement, and the low [Ba/Fe] ratios (with only the exceptional case of S3), suggest minor contribution from AGB-mass transfer (if any). 

 The values obtained from our HR analysis for [Mg/Fe], [Si/Fe], and [Ti/Fe] can now be compared with those reported in Table~\ref{Tab:stars} obtained from the RAVE spectra. In most of the cases the discrepancies are above the quoted error bars, and it is probably due to the combination of the lower resolution and shorter spectral coverage of RAVE spectra, that leads to undetected line blends and the presence of very few lines per element. The [$\alpha$/Fe] ratios coming from high-resolution UVES spectra show a large variation. Enhancements for S2 and S4 seem systematically larger than the ones of S1 and S3. 

In Fig.~\ref{Fig:datalit} the atmospheric parameters in this work (from RAVE and UVES spectra) are compared with the literature values presented in RAVE-DR5 (calibrated values), RAVE-on (\citealt{Casey2017}, where the stellar parameters were obtained by using a data-driven approach). It is worth noticing that the RAVE-on catalogue misplaced these red giants in metallicity and/or gravity. This misclassification might be due to the training sample adopted in \citet{Casey2017}, consisting mostly of APOGEE red giants, that are mostly metal rich. In Fig.~\ref{Fig:datalit} is visible also that for the star 201359581 the temperature obtained with the \cite{Valentini2017} method is $\sim$350 K higher than the one measured from the high-resolution spectrum. This is a consequence of the fact that the starting \Teff adopted was erroneous. For stars S2, S3 and S4, there is a good agreement between the temperatures estimated from the RAVE and high-resolution analysis spectra, upon the use of the seismic gravity. The agreement is also seen in metallicity, where the most discrepant case, S4, is our most metal-poor star for which the non-local thermodynamic equilibrium (NLTE) corrections are more important (we took into account NLTE effects, when analysing UVES spectra). Two important results can be extracted from Fig.~\ref{Fig:datalit}: {\it i}) by combining the RAVE spectra with  seismic gravities it is possible to reach precise stellar parameters, similar to what is obtained from high-resolution spectra (see the agreement between the black dots (UVES) and red points (RAVE) for 3 out of the 4 stars); {\it ii}) the high-resolution analysis has confirmed that one of the stars has metallicity [Fe/H] $<-$ 2. The difficulty in determining the metallicity of such metal-poor objects from moderate resolution spectra covering a rather short wavelength range, not having the extra seismic information, is clearly illustrated by the discrepant metallicities found by RAVE DR5 and RAVE-on, versus the good agreement with the value published in Valentini et al. (2017) upon the use of K2 information, where the temperatures and gravities are consistent.

\section{Mass and age determination}
\label{Sect:massage}
Mass determinations have been performed using two different methods: {\it i)} a direct method, using scaling relations, and {\it ii)} a Bayesian fitting using the PARAM code \citep{Rodrigues2017}. Masses derived using scaling relation differ from the ones from PARAM (see discussion in Rodrigues et al. 2017). We now illustrate this difference for the case of our four metal-poor stars.
The resulting masses from the two methods are summarized in Table~\ref{Tab:seismres}.

\subsection*{Mass estimate using scaling relations:}

For our computations using the scaling relations we adopt as input \dnu~ and \numax~  from the COR pipeline and the \Teff measured from the UVES spectra. The scaling relations are in the form:
\begin{eqnarray}
\label{Eq:scaling}
\frac{M}{M_\odot} &\simeq& \left(\frac{\nu_{\rm max}}{\nu_{\rm max, \odot}}\right)^{3}\left(\frac{\Delta\nu}{\Delta\nu_{\odot}}\right)^{-4}\left(\frac{T_{\rm eff}}{T_{\rm eff, \odot}}\right)^{3/2}\\
\frac{R}{R_\odot} &\simeq& \left(\frac{\nu_{\rm max}}{\nu_{\rm max, \odot}}\right)\left(\frac{\Delta\nu}{\Delta\nu_{\odot}}\right)^{-2}\left(\frac{T_{\rm eff}}{T_{\rm eff, \odot}}\right)^{1/2}
\end{eqnarray}
were the solar values adopted are the same ones listed in Section~\ref{Sect:data}, and \Dnu=135.1 $\mu$Hz.

The uncertainties on the masses and radii are  calculated using propagation of uncertainties, under the assumption of uncorrelated errors. .

\subsection*{Mass estimate using PARAM:}

For deriving ages and masses via Bayesian inference we adopted the latest version of the PARAM code. The new version of the code uses \dnu that has been computed along MESA evolutionary tracks, plus \Numax computed using the scaling relation. The following modifications were implemented with respect to the version described in \citet{Rodrigues2017},  namely: {\it i}) we extended the grid towards the metal poor end, down to [Fe/H]=$-$3 dex, by calculating evolutionary tracks for \FeH = $-$2.0 and $-$3.0 dex, with He enrichment computed according \citet{Rodrigues2017}; and {\it ii}) we took $\alpha$-elements enrichment into account, by converting the observed chemical composition into a solar-scaled equivalent metallicity. {\rm We investigated the solutions provided by PARAM when setting an upper limit to the age at 14~Gyr and without age upper limit (the latter helps in understanding the shape of the PDF of mass and age).}

\subsection{Mass-loss and alpha-enhanced tracks}

\begin{figure}
   \centering
   \includegraphics[width=0.9\columnwidth]{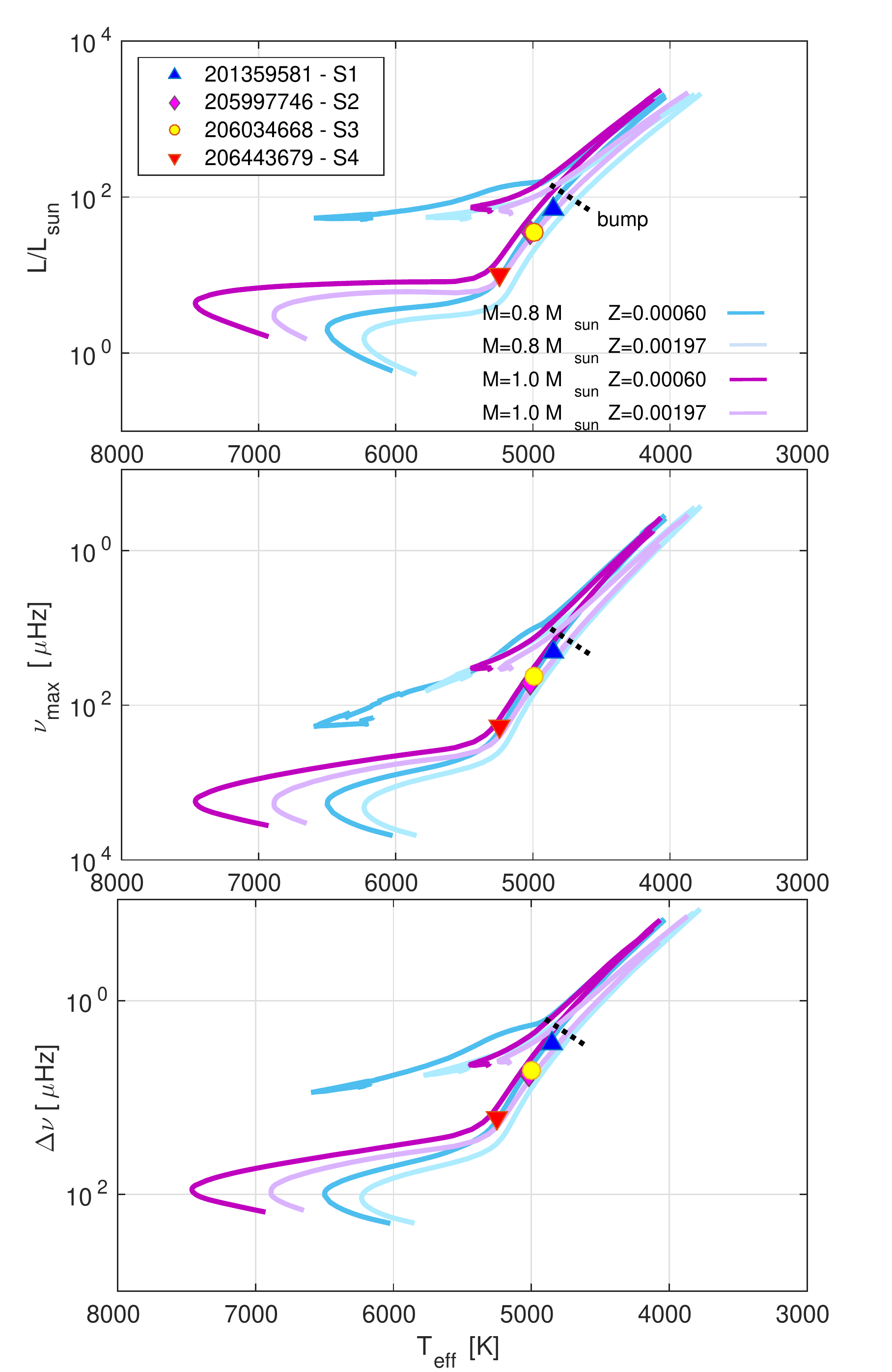}
   \caption{Top panel: Position in the temperature - luminosity diagram of the four RAVE stars of this work (nomenclature following Table 1). Evolutionary tracks at masses M= 0.8 and 1.0 \Msun, at two different metallicities (Z=0.00060 and 0.00197) are plotted. Middle panel: Position in the temperature - \Numax diagram of the four RAVE stars of this work, same tracks as top panel. Bottom panel: Position in the temperature - \Deltanu diagram of the four RAVE stars of this work, same tracks as top panel. Error bars of the plotted quantities are of the size of the points.}
              \label{Fig:tracks}%
    \end{figure}

PARAM provides also an estimate for stellar distance and luminosity, $L$ (listed in Table~\ref{Tab:seismgaia}). The luminosities provided by PARAM were used to construct Fig.~\ref{Fig:tracks}, where we placed our stars in the temperature-luminosity diagram. The figure shows a set of MESA evolutionary tracks for masses 0.8 and 1.0 \Msun, at two different metallicities Z=0.00060 and Z=0.00197. In the same figure the four stars are also plotted, together with the track, in the \numax-\teff (middle panel) and \dnu-\Teff planes. The stars of our sample are most likely low-luminosity RGB stars which are not expected to undergo significant mass loss. The evolutionary state of star S1 (201359581), on the other hand, is more uncertain, since it is locate close to the RGB bump (dashed line), following also Fig.~1 of \citet{Khan2018}, it can be core-He burning, RGB, early AGB. The evolutionary status of this star becomes relevant when it comes to discussing the reliability of age estimates, since stars in the red-clump or early AGB phases suffer of significant mass loss, that hampers the mass (and hence age) determination. Finally, since our stars are well located below the bump (with a flag on S1 that is an borderline case), we consider their abundances not affected by extra-mixing process that happens at the bump and early AGB stage.

Because the adopted MESA stellar tracks in \citet{Rodrigues2017} assume the \citet{Grevesse1993} solar mixture for the metals, we adopt the $\alpha$-enhancement correction to convert [Fe/H] into [M/H] by using the formula from \citet{Salaris1993}, updated using the relative mass fraction of elements from OPAL tables \footnote{\url{https://opalopacity.llnl.gov/pub/opal/type1data/GN93/ascii/GN93hz}} :

\begin{equation}
\label{Eq:salaris}
{\rm [M/H]}^{\rm chem}= {\rm [Fe/H]}+ \log{\left( C \times 10^{[\alpha/{\rm Fe}]}+(1-C)\right)}
\end{equation}
where $C$=0.684.

This is a necessary step, given that all our stars are $\alpha$-enhanced. We tested the effectiveness of this assumption by comparing two PARSEC track sets (from MS to RGB tip), which are also provided for $\alpha$-enhanced cases. In Appendix~\ref{AlphaJust}) we compare one track computed for [$\alpha$/Fe]=+0.4 dex and \FeH=$-$2.15, and one not alpha enhanced, but with the corresponding metallicity following Equation~\ref{Eq:salaris} (\FeH =$-$1.86). The test shows that the deviation in age between the tracks has its maximum at the RGB tip, in the mass regime of our stars. This deviation is in the order of 1-2\%, a smaller effect respect to the typical age uncertainty.
We derived mass and ages by adopting first the atmospheric parameters derived from RAVE spectra and then for the atmospheric parameters obtained from UVES spectra. We also computed mass and ages using the different seismic inputs discussed in Section~\ref{Sect:data} (COR and BM$\_$N). This strategy allows us to see the impact of different precision in the atmospheric parameters and seismic parameters. Results are summarised in Appendix Table~\ref{Tab:PARAMpipelines}. Results obtained with the high-resolution input for temperature, metallicity, and an averaged [$\alpha$/Fe] (computed as ([Mg/Fe]+[Si/Fe]+[Ca/Fe])/3) are in Fig.~\ref{Fig:BM_vs_BMN} and in Appendix Fig.~\ref{Fig:compAgeLim}. In these figures it is visible that the PDF of masses and ages obtained with the seismic values with BM\_N seismic values are broad and, in the case of 205997746, double peaked. This is a consequence of the inflated error in BM\_N, caused by blindly combining all the spectroscopic pipelines. This shows that, when dealing with a detailed analysis of individual stars, a star-by-star approach for testing the performances of each seismic pipeline is a necessary step for increasing the precision of mass and age determination.  
\begin{table*}
\centering
\begingroup
\renewcommand{\arraystretch}{1.2} 
\caption{Seismic mass and radius calculated using scaling relations (\Teff measured from UVES spectra), and mass, radius, and age derived using PARAM, for the 4 metal-poor RAVE stars in K2 Campaigns 1 and 3. The last column lists the stellar radius provided by Gaia DR2.}
\label{Tab:seismres}
\centering          
\begin{tabular}{lcccccccc}     
\hline\hline       
ID & M$_{\rm scaling}$ & R$_{\rm scaling}$ & M$_{PARAM}$ & R$_{PARAM}$ & Age$_{PARAM}$ & R$_{\rm Gaia DR2}$   \\
this work &  [\msun] & [\rsun]  & [\msun] & [\rsun] & [Gyr]  \\  \hline
S1 & 1.18$\pm$0.16 & 14.04$\pm$0.16 & 0.96$^{+0.11} _{-0.08}$ &12.65$^{+0.60} _{-0.57}$  &  7.42$^{+2.12} _{-2.68}$   &16.84$^{+0.58} _{-0.30}$\\
S2 & 1.12$\pm$0.06 &  8.53$\pm$0.07 & 0.99$^{+0.08} _{-0.07}$ & 7.96$^{+0.24} _{-0.26}$  &  7.76$^{+1.24} _{-2.74}$   & 8.19$^{+0.08} _{-0.10}$\\
S3 & 0.87$\pm$0.10 &  8.30$\pm$0.19 & 0.78$^{+0.11} _{-0.10}$ & 7.72$^{+0.39} _{-0.41}$  & 13.01$^{+12.99} _{-3.15}$  & 9.24$^{+1.43} _{-0.43}$\\
S4 & 1.01$\pm$0.12 &  4.15$\pm$0.15 & 0.87$^{+0.08} _{-0.08}$ & 3.90$^{+0.14} _{-0.12}$  &  9.58$^{+3.68} _{-2.57}$  & 3.97$^{+0.20} _{-0.19}$\\
\hline
\end{tabular}
\endgroup
\end{table*}

   \begin{figure*}
   \centering
   \includegraphics[width=2\columnwidth]{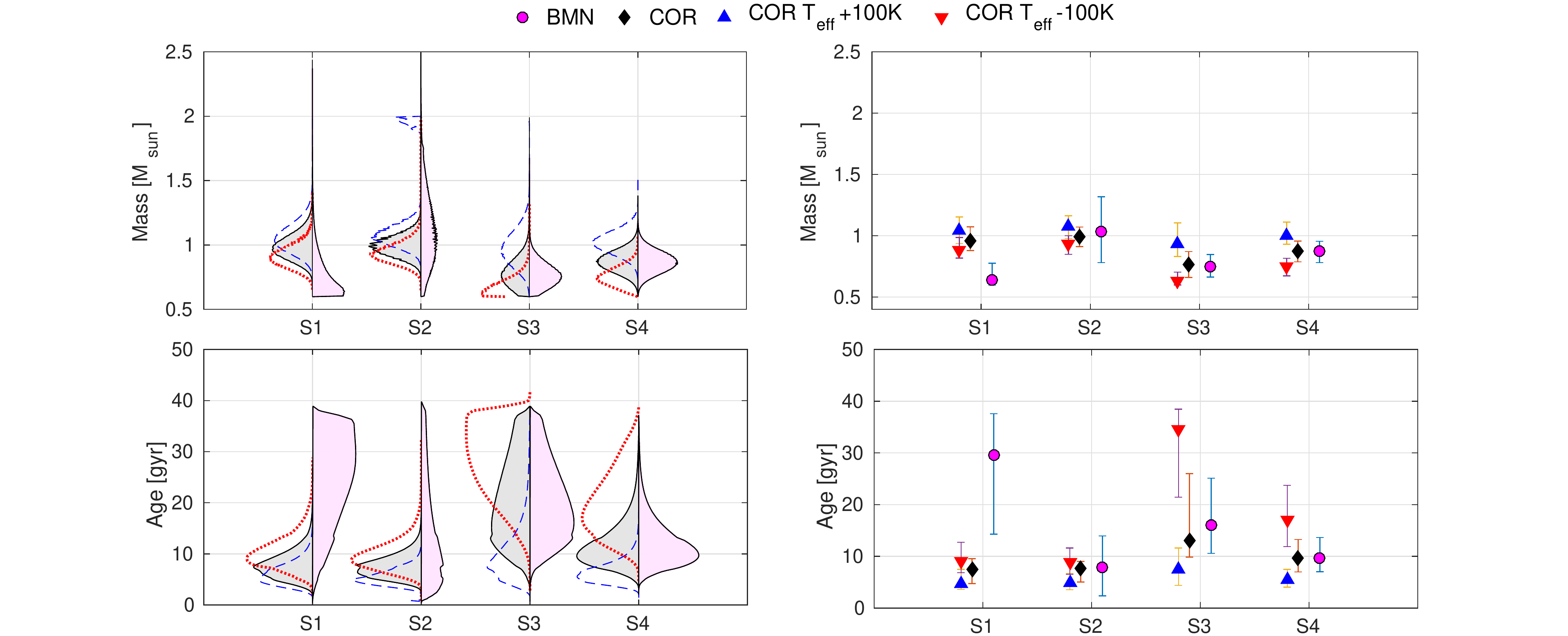}
   \caption{Left column: violin plot of the PDFs of mass (top) and age (bottom). The right magenta shaded PDF is derived using the seismic parameters from BM\_N seismic set of parameters, with the new errors that take into account dispersion between pipelines, the PDFs on the left of the violin are calculated using seismic parameters from COR pipeline (black line, gray shaded) and varying the \Teff of +100 K (dashed blue line) or $-$100 K (dotted red line). Right column: modes and 68 percentile errorbar of masses (top) and ages (bottom) of the 4 stars of this work. Magenta points are values computed using BM$\_$N seismic values, black diamonds are the values derived using COR seismic values, and red and blue triangles are values obtained using COR seismic values and varying the \Teff of $-$100 and +100 K respectively.}
   \label{Fig:BM_vs_BMN}
    \end{figure*}

Our adopted final values of stellar mass and radius, derived using COR seismic input and UVES spectra, are shown in Tab.~\ref{Tab:seismres}, where we also show, for comparison, the results obtained directly from the scaling relations. The mass and ages of PARAM are obtained adopting a mass-loss value derived from \citet{Reimers1975} law with an efficiency parameter of $\eta$=0.2. We adopted this value since it is in agreement to what measured in \cite{Miglio2012} by comparing the asteroseismic masses of Red Clump stars and Red Giants in the old open clusters NGC6791 and NGC6819. The error associated to the mode value of radius, mass and age derived using PARAM is calculated as the shortest credible interval with 68 per cent of the probability density function (PDF).
Masses derived with scaling relations (Eq.~\ref{Eq:scaling}) are larger than those derived using PARAM by circa 30\%. This is due to the correction needed to \Dnu (see \cite{Miglio2016}), that leads to a more accurate mass estimation for red giants. In PARAM this correction is not necessary. The code can, in fact, derive the theoretical \Dnu directly by interpolation, since this quantity has been estimated along each evolutionary track.
 
\subsection{Using Luminosities from Gaia DR2 to further constrain PARAM}

In the work of \citet{Rodrigues2017} the adoption of the intrinsic stellar luminosity, $L$, derived using Gaia parallaxes, leads to a significant improvement into the mass and age determination (from an error of 5\% in mass and 19\% in age to 3\% and 10\% respectively). These estimates were based on high-quality {\it Kepler} seismic data and very precise atmospheric parameters. In addition, the uncertainties on luminosity were assumed to be  3\%, from Gaia end-of-the-mission performances. Gaia DR2 does not still reach this precision and offsets in $\varpi$ have to be taken into account. Nevertheless we calculated mass, radius and age using the additional information on $L$, calculated from parallax and find out the shape of the PDFs were affected, suggesting some tension with the input luminosities.

Instead of using the luminosities tabulated in Gaia DR2, we considered the weighted mean of the $L$ calculated from K$s$, I, and V magnitudes, considering BC provided by \citet{BC2014} and \citet{BC2018} \footnote{Codes available at \url{https://github.com/casaluca/bolometric-corrections}} and the reddening derived from \citet{Schlegel1998} maps. Errors on $L$ were calculated via error propagation, with the error on BC calculated via Monte-Carlo simulation of 100 points for each star. Luminosities are listed in Table~\ref{Tab:seismgaia} and show $\sim$ 15\%  uncertainties, and not the  3\%  end of mission expectation. We thus opted for not using luminosities as an extra constraint in our calculations of mass and radius.

\begin{table*}
\centering  
\begingroup
\renewcommand{\arraystretch}{1.2} 
\caption{Seismic distances and luminosities calculated using scaling relations, PARAM, distances obtained from Gaia parallaxes (both using the classical 1/$\varpi$ and \cite{BJ2018}), distance calculated using StarHorse \citep{Queiroz2018}, and luminosity provided by Gaia DR2. For calculating $L$ from $\varpi_{\rm Gaia DR2}$ we used the bolometric corrections of \citet{BC2018}. Stars are identified using the nomenclature in Table~\ref{Tab:stars}.}
\label{Tab:seismgaia}
        
\begin{tabular}{lccccccccccc}     
\hline\hline       
ID    & Dist$_{\rm scaling}$ & Dist$_{PARAM}$       &Dist$_{\rm \varpi~Gaia DR2}$& Dist$_{SH}$         &Dist$_{\rm \varpi~Gaia DR2~B-J}$ &L$_{PARAM}$            &L$_{\varpi B-J}$   & L$_{\rm Gaia DR2}$   \\
      & [pc]          & [pc]                        & [pc]                       & [pc]                &[pc]                             &[L$\odot$]             & [L$\odot$]        & [L$\odot$]   \\  \hline
S1 *  & 1634 $\pm$322 & 1536 $^{+66} _{-49}$  & 1907$_{-274} ^{+370}$      & 1665$_{-69} ^{+30}$ & 2194$_{-434} ^{+1389}$          &89.7$^{+9.5} _{-9.2}$  & 91.0 $\pm$ 17.2   & 158.1 $^{+36.9} _{-36.9}$ \\
S2    & 2094 $\pm$546 & 1879 $^{+59} _{-73}$  & 2099$_{-214} ^{+270}$      & 1961$_{-59} ^{+69}$ & 1945$_{-369} ^{+231}$           &38.2$^{+2.8} _{-3.0}$  & 32.3 $\pm$ 5.8    & 37.3$^{+5.7} _{-5.7}$\\
S3    & 1378 $\pm$355 & 1446 $^{+59} _{-56}$  & 1659$_{-99} ^{+113}$       & 1485$_{-10} ^{+3}$  & 1579$_{-232} ^{+104}$           &35.7$^{+4.3} _{-4.1}$  & 51.0 $\pm$ 7.6    & 49.6$^{+4.7} _{-4.7}$\\
S4    & 675  $\pm$187 & 706  $^{+22} _{-13}$  & 725$_{-22} ^{+24}$         & 798$_{-12} ^{+28}$  & 710$_{-21} ^{+23}$              &11.1$^{+0.9} _{-0.9}$  & 11.2 $\pm$ 1.4    & 9.7$^{+0.5} _{-0.4}$\\
\hline
\end{tabular}
\begin{tablenotes}
      \small
      \item (*) Gaia DR2 values for S1 (201359581) are flagged for duplicity and astrometric noise (see Table~\ref{Tab:data}). For this reason distance and luminosity obtained using Gaia $\varpi$ are not reliable.   
    \end{tablenotes}
\endgroup
\end{table*}


\subsection{Uncertainties}


For better understanding the systematics that may affect the age determination using PARAM we performed several tests under different assumptions:
\begin{itemize}
\item We determined ages and masses for each set of seismic parameters provided by different pipelines.
\item We used atmospheric parameters from RAVE and UVES spectra.
\item  For each set of seismic parameters, when using atmospheric parameters derived from RAVE spectra, we considered five different [$\alpha$/Fe] abundances: 0.0, 0.1, 0.2, 0.3, 0.4 dex. Since the low resolution and the limited wavelength interval of RAVE may affect the measured alpha content of the stars, we wanted to quantify the impact of a erroneous [$\alpha$/Fe]. 
\item Two different mass loss efficiency parameters were considered, $\eta=$0.2 and 0.4. This test has been performed for each set of seismic data adopted.
\item We varied the \Teff of $\pm$ 100K, this shift is for simulating the effect of a difference in temperature that may exist between different methods for measuring it.
\item We tested the impact of the precision on \Teff, by adopting as input error a value two times the spectroscopic value. Resulting masses and ages are listed in Appendix Table~\ref{Tab:PARAMpipelines}, in the rows labeled as ''COR$_2\sigma{\rm Teff}$''.
\item {\rm We tested the introduction of an upper limit on the age. Resulting masses and ages are listed in Appendix Table~\ref{Tab:PARAMpipelines}, in the rows labeled as ''COR$_{\rm agelim}$''.}
\end{itemize}

It is worth to remember that the effects of these tests depend on the position of the star on the HR diagram, and on its evolutionary stage. Each locus of the HR diagram is populated by different tracks and with different levels of crowdedness. 

The variation on $\alpha$ content has no significant effect, providing a mass spread on average of 0.01 \Msun and of 0.3 Gyr in age (see Appendix Figs.~\ref{Fig:agesBenoit} and \ref{Fig:agesGuy}). As a general behaviour, when the $\alpha$ enrichment increases the mass slightly decreases and the age increases. 

The underestimation of \Teff of 100~K leads to a variation in mass and age on average of $-$10\% and $+$30\% respectively. As expected, when temperature increases the mass increases and the age decreases, the contrary happens when the temperature decreases. This effect is more visible for the most metal poor and hottest stars.

The adoption of an inflated error on \Teff, two times the nominal spectroscopic error, lead to no sensible change in the mass and age determination. When adopting a seismically determined \Teff, we are taking advantage of using a \Teff that is consistent with the seismic parameters themselves. In the case of an inaccurate \Teff, as for S1 using RAVE spectra, the solution is misleading and PARAM shows tensions in the posterior PDFs (see Appendix Fig.~\ref{Fig:tensions}).

The adoption of a mass-loss parameter $\eta$=0.4 leads to a mass increase of only 2\% and an age reduction of 4\% in mass and age respect to the values derived with $\eta$=0.2  (see Appendix Figs.~\ref{Fig:agesBenoit} and \ref{Fig:agesGuy}). As discussed in \citet {Anders2016} and \citet{Casagrande2016}, the effect of mass loss is more significant for red clump stars than for RGB stars. Three of the four stars studied here are consistent with the RGB classification (see Fig.~\ref{Fig:tracks}), so our results appear consistent with their findings.

Setting a uniform prior on age with an upper limit has a consequence on the shapes of the PDFs of ages. This is the reason why, in some cases, for the oldest stars of the sample, the PDF of the age appears truncated at the upper limit, as visible in Appendix Fig.~\ref{Fig:compAgeLim}. {Although not considering an upper limit on the age at 14 Gyr would not represent the information that we have about the age of the Universe, removing the age limit allows the PDF to extend to older ages, so we can better understand its shape and therefore the goodness of the age determination (e.g. multiple peaked PDF). As visible in Appendix Fig.~\ref{Fig:compAgeLim} and listed in Appendix Table~\ref{Tab:PARAMpipelines}, the removal or adoption of an upper age limit has little consequence on the resulting mass and ages.

\subsection{Masses and ages for two previously studied samples and comparison with our sample}

We compare the masses of our stars with the masses previously determined in the literature for metal-poor field giants in the APOKASC sample \citep{Epstein2014} and for giants in the globular cluster M4 \citep{Miglio2016} also using asteroseismic information. We also recomputed masses and ages for the two literature samples using PARAM in the same set-up used for the RAVE metal-poor stars analysed in this work.  
\subsubsection*{The APOKASC metal-poor giants}
For the APOKASC targets of \citet{Epstein2014} we adopted atmospheric parameters and their uncertainties from APOGEE-DR14 \citep{APODR14} together with \Dnu and \Numax obtained by the COR pipeline from \kepler light-curves (this choice is needed for granting homogeneity in our sample). The atmospheric parameters of APOGEE-DR14 differ from those adopted by \cite{Epstein2014}, since that work used previous ASPCAP releases. The input parameters we used in PARAM are given in Table~\ref{Tab:PARAM_Epstein}, where the metallicities [M/H] are computed with Eq.~\ref{Eq:salaris} to take into account the [$\alpha$/Fe]-enhancement.
The PARAM code provided mass and age for each star of the \cite{Epstein2014} work. We did not considered the results for star E14-S5, since the resulting a-posteriori \teff, \numax, and \Dnu were not in agreement with the input values (see Appendix~\ref{app:tensions} for an example), indicating the presence of erroneous input parameters. The masses we obtain are now smaller with respect to the original values of Epstein et al. (2014) who reported masses obtained using scaling relations. The new masses are also in agreement with the masses we obtained for the four RAVE stars (see Fig.~\ref{Fig:metal-poor} upper panel). The differences in masses between the \cite{Epstein2014} estimates and ours are consistent with the fact that the scaling relation masses are systematically larger than the ones computed by PARAM for RGB stars (as previously discussed, see Table~\ref{Tab:seismres}). The two samples together provide a better coverage of the metal-poor \FeH regime.
Masses of the \citet{Epstein2014} sample have been already recomputed by \citet{Sharma2016}, \citet{Pinsonneault2018}, and \citet{Yu2018}, taking into account \Dnu corrections derived from stellar models. In Fig.\ref{Fig:comparison}, we compared the masses of the \kepler metal-poor stars of \citet{Epstein2014}, with those computed in this work, \citet{Pinsonneault2018}, and \citet{Yu2018} ones (the lattest considerig the value corresponding to their evolutionary status). Masses agree within errors. On the other hand, \citet{Yu2018} masses are in general always larger than those we computed in this work, resulting into smaller ages. This might be the result of the different set of atmospheric parameters adopted by the authors.

   \begin{figure}
   \centering
   \includegraphics[width=1\columnwidth]{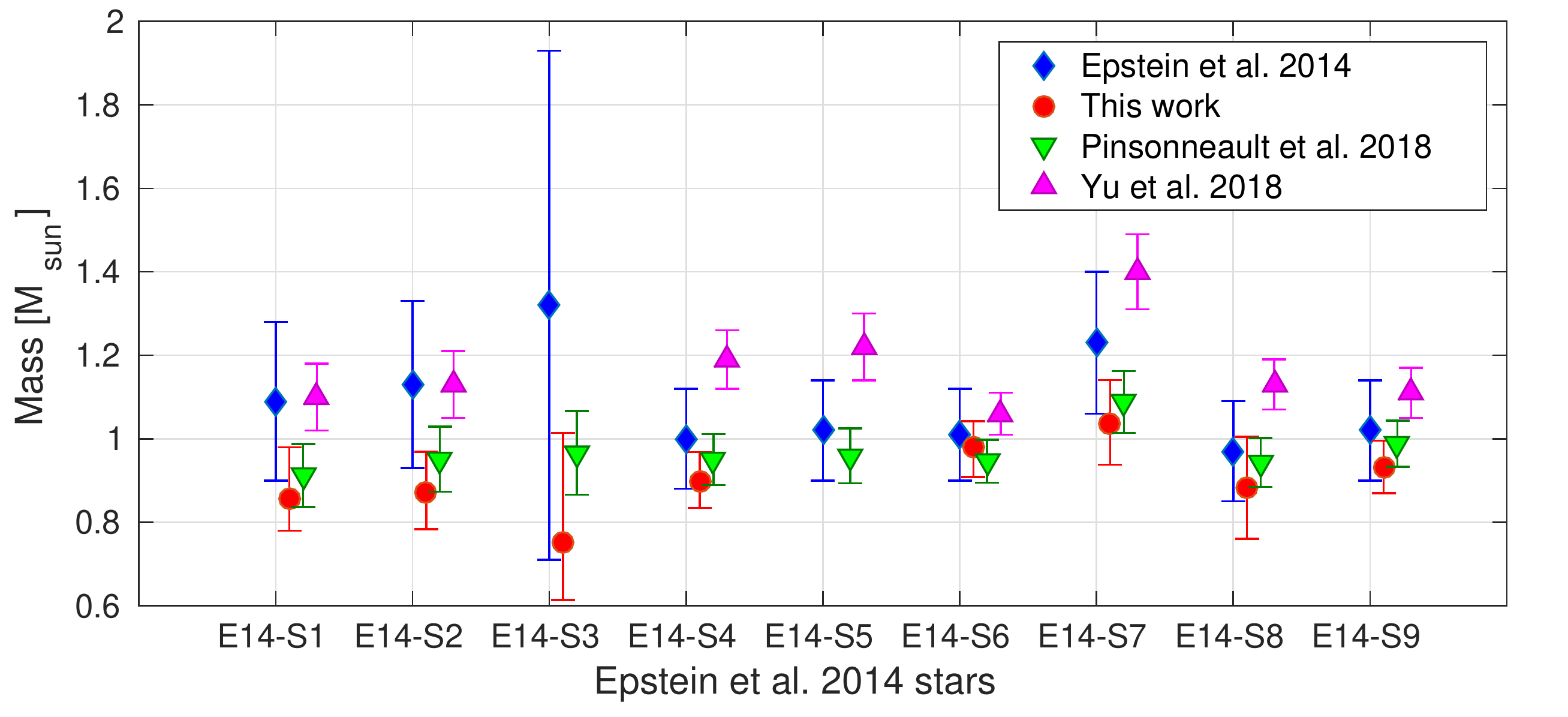}
   \caption{Mass and age comparison for the 9 stars presented in \citet{Epstein2014} (blue diamonds) with the values derived in this work (red circles), in \citet{Pinsonneault2018} (green triangles) and \citet{Yu2018} (magenta triangles). Stars are indexed as in Table~\ref{Tab:PARAM_Epstein}}
              \label{Fig:comparison}%
    \end{figure}
\subsubsection*{The red giants in the M4 globular cluster}
We also provide a similar comparison for seven M4 stars previously studied by \cite{Miglio2016} for which K2 seismic information were available. This sample is an ideal benchmark for testing our method, since for globular clusters a reliable and precise age can be measured. In this case the temperature was obtained from (B--V) colour (corrected) as in Casagrande \& VandenBerg (2014), assuming a temperature uncertainty of 100 K. The input parameters adopted in this case are summarised in Table~\ref{Tab:PARAM_M4}. Our masses and ages determinations are consistent with the original values of \cite{Miglio2016} who, despite  of using the scaling relations, took the necessary correction for RGB stars into account. With the exception of one outlier (M4-S6), the stars provide an age for the globular cluster of $\sim$11.01$\pm$2.67 Gyr (derived using as the weighted mean based on the mean error, $\sim$11.80$\pm$2.58 Gyr when considering all the stars), in agreement with the age measured from isochrone fitting of 13 Gyr \cite[or with the 12.1$\pm$0.9 Gyr,][age measured from the white dwarf cooling sequence]{Hansen2004}. The PDF of mass and age for the individual stars of M4 are plotted in Appendix Fig.~\ref{Fig:M4pdfs} and compared with the literature values.

Figure~\ref{Fig:metal-poor} summarises the ages and masses obtained in the present work for the three datasets (i.e. four RAVE, eight APOKASC, and seven M4 stars). In this Figure we have plotted the ages and masses estimates for the four stars obtained by using the COR pipeline seismic inputs consistent with the seismic inputs in Tables~\ref{Tab:PARAM_Epstein} and \ref{Tab:PARAM_M4} of the other two samples analysed here. All the 19 stars plotted in the Figure (filled symbols) are compatible with masses below one solar mass (upper panel). Most of these halo objects are consistent with being very old, and none is younger than $\sim$7 Gyr. Moreover, our results show that it is possible to estimate ages for metal-poor giants with seismic information, not only from \emph{Kepler}, but also from the K2 less precise light curves.

   \begin{figure*}
   \centering
   \includegraphics[width=2\columnwidth]{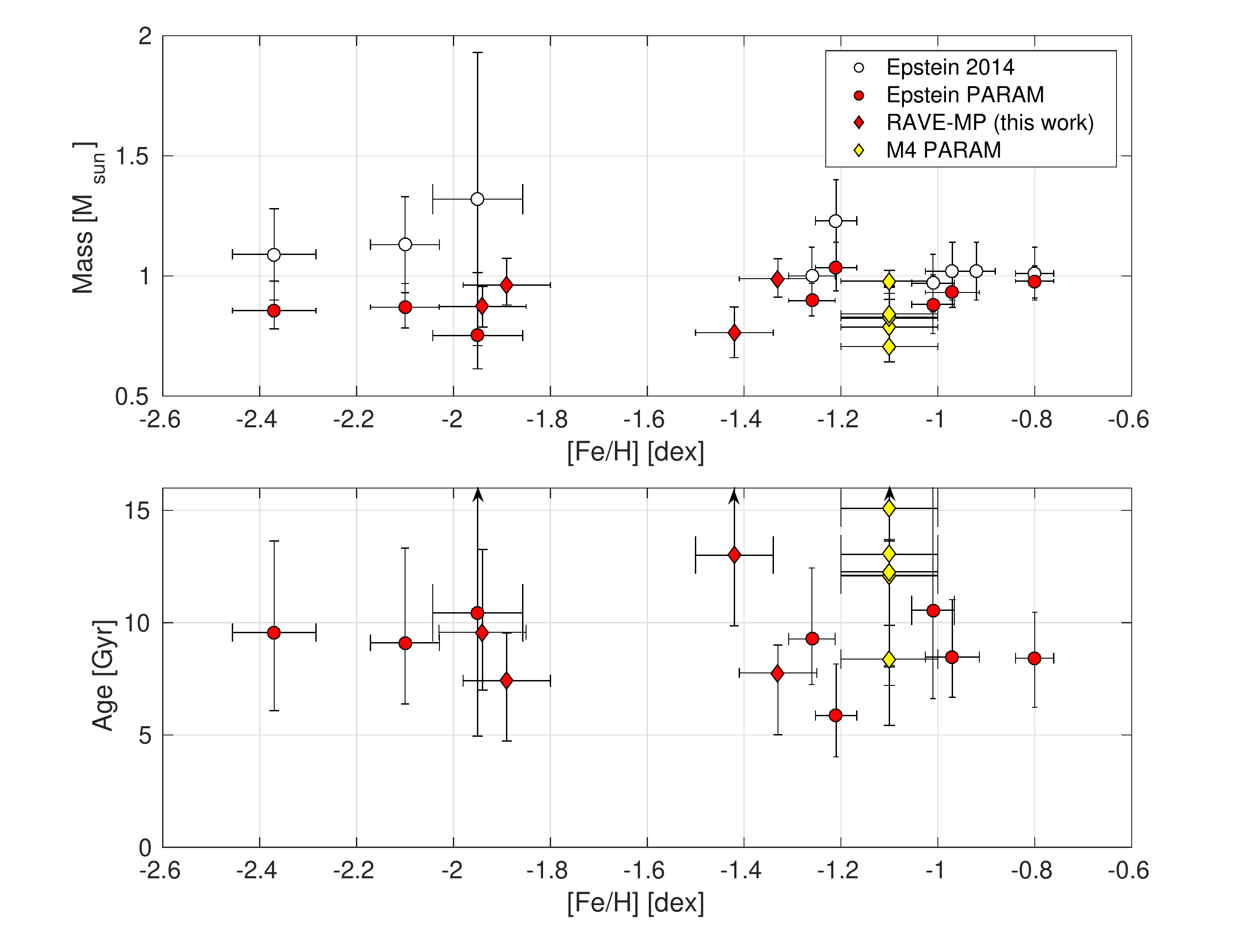}
   \caption{Mass and ages of red giant branch stars in the metal-poor regime for a) four metal-poor RAVE stars with K2 seismic oscillations presented in this work (red diamonds) ; b) nine APOKASC the objects from \cite{Epstein2014} (original values as empty black circles; our new determination using PARAM and APOGEE-DR14 atmospheric parameters and abundances are shown as filled red circles), and c) seven stars in M4 from  \cite{Miglio2016} recomputed with PARAM in this work (yellow triangles).}
              \label{Fig:metal-poor}%
    \end{figure*}


\begin{table*}
\begingroup
\renewcommand{\arraystretch}{1.2} 
\caption{Input atmospheric parameters for the \cite{Epstein2014} stars and PARAM results of mass and age. [M/H] was computed using the Equation~\ref{Eq:salaris}. Atmospheric parameters and \FeH come from APOGEE DR14. Seismic \Dnu and \Numax are derived using COR pipeline (source: APOKASC catalogue 4.4.2).}
\label{Tab:PARAM_Epstein}
\centering
\begin{tabular}{llrrrrrrr}
\hline \hline
ID  & KIC ID  & \Dnu   &\numax & [M/H]  &\Teff &  [Fe/H] &Age & Mass  \\ 
    &         &     [$\mu$Hz] & [$\mu$Hz] & [dex]  & [K]   & [dex] & [Gyr] &[\Msun]  \\ \hline
E14-S1 &  7191496 & 2.45$\pm$0.05 & 16.10$\pm$0.28&$-$1.97$\pm$0.65 & 4912$\pm$87 &$-$2.37$\pm$0.11 &9.57$_{-3.48}^{+4.07}$ & 0.86$_{-0.08}^{+0.12}$\\ 
E14-S2 & 12017985 & 2.64$\pm$0.05 & 17.80$\pm$0.32&$-$1.85$\pm$0.44 & 4908$\pm$86 &$-$2.10$\pm$0.10 &9.11$_{-2.73}^{+4.21}$ & 0.87$_{-0.09}^{+0.10}$\\ 
E14-S3 &  8017159 & 0.69$\pm$0.05 &  3.10$\pm$0.14&$-$1.72$\pm$0.39 & 4629$\pm$72 &$-$1.95$\pm$0.07 &10.44$_{-5.49}^{+19.57}$& 0.75$_{-0.14}^{+0.26}$\\ 
E14-S4 & 11563791 & 5.06$\pm$0.05 & 42.50$\pm$0.72&$-$0.99$\pm$0.45 & 4929$\pm$87 &$-$1.25$\pm$0.06 &9.28$_{-2.03}^{+3.16}$ & 0.89$_{-0.06}^{+0.07}$\\ 
E14-S5*& 11181828 & 4.14$\pm$0.05 & 33.30$\pm$0.57&$-$0.77$\pm$0.27 & 4790$\pm$83 &$-$0.92$\pm$0.05 &3.37$_{-0.22}^{+6.84}$ & 1.19$_{-0.30}^{+0.02}$\\ 
E14-S6 &  5858947 &14.54$\pm$0.05 &169.30$\pm$2.88&$-$0.64$\pm$0.30 & 5002$\pm$87 &$-$0.81$\pm$0.05 &8.42$_{-2.18}^{+2.05}$ & 0.98$_{-0.07}^{+0.06}$\\
E14-S7 & 7019157  & 3.49$\pm$0.05 &  27.5$\pm$0.48&$-$0.99$\pm$0.37 & 4820$\pm$90 &$-$1.21$\pm$0.05 &5.87$_{-1.84}^{+2.29}$ & 1.03$_{-0.10}^{+0.11}$\\ 
E14-S8 & 4345370  & 4.09$\pm$0.05 & 32.40$\pm$0.56&$-$0.83$\pm$0.31 & 4791$\pm$77 &$-$1.01$\pm$0.05 &10.56$_{-3.94}^{+8.71}$ & 0.88$_{-0.12}^{+0.12}$\\
E14-S9 & 7265189  & 8.57$\pm$0.05 & 85.10$\pm$1.45&$-$0.85$\pm$0.23 & 4996$\pm$87 &$-$0.97$\pm$0.05 &8.48$_{-1.79}^{+2.55}$ & 0.93$_{-0.06}^{+0.06}$\\ 
       \hline
\end{tabular}
\begin{tablenotes}
 \small
      \item (*) {\bf Star E4-S5 presented tensions between input parameters and output parameters (similar to the case presented in Appendix~\ref{app:tensions}. For this reason, we disregarded this result, even if we report the result in this table.}).   
    \end{tablenotes}
  \endgroup  
\end{table*}

\begin{table*}
\begingroup
\renewcommand{\arraystretch}{1.2} 
\caption{Input atmospheric parameters for the stars analysed in the M4 globular cluster \cite{Miglio2016} adopted in PARAM and resulting mass and age. [M/H] has been computed using the Equation~\ref{Eq:salaris} {\rm (\FeH =$-$1.1 dex and [$\alpha$/Fe]=+0.4 dex)}.}
\label{Tab:PARAM_M4}
\centering
\begin{tabular}{lrrrrrrrrrr}
\hline \hline
ID&   \dnu   &  $\sigma$\Dnu &  \numax &   $\sigma$\numax&    [M/H]    & $\sigma$[M/H]&   \teff & $\sigma$\teff   &    Age & Mass   \\ 
 &     [$\mu$Hz] &  [$\mu$Hz] & [$\mu$Hz] & [$\mu$Hz] & [dex] & [dex] & [K]  & [K] & [Gyr]  &[\Msun]  \\ \hline
M4-S1&   1.83 &   0.02 &   11.1 &   0.4   &   -0.80 &  0.13 &   4585 &  100 &  15.09$_{-5.20} ^{+10.63}$ & 0.79 $_{-0.09}^{+0.11}$\\ 
M4-S2&   2.55 &   0.04 &   17.2 &    0.7  &   -0.80 &  0.13 &   4715 &  100 &  12.11$_{-4.90} ^{+11.05}$ & 0.83 $_{-0.12}^{+0.14}$\\
M4-S3&   2.62 &   0.04 &   17.7 &    0.7  &   -0.80 &  0.13 &   4710 &  100 &  13.05$_{-5.01} ^{+11.75}$ & 0.83 $_{-0.13}^{+0.11}$\\
M4-S4&   2.64 &   0.02 &   18.5 &    0.7  &   -0.80 &  0.13 &   4715 &  100 &   8.38$_{-2.95} ^{+5.33}$  & 0.98 $_{-0.16}^{+0.08}$\\
M4-S5&   4.14 &   0.02 &   32.5 &    1.3  &   -0.80 &  0.13 &   4847 &  100 &  12.07$_{-3.97} ^{+10.05}$ & 0.82 $_{-0.11}^{+0.12}$\\
M4-S6&   4.30 &   0.02 &   32.9 &    1.3  &   -0.80 &  0.13 &   4842 &  100 &  22.79$_{-9.15} ^{+10.69}$ & 0.71 $_{-0.06}^{+0.13}$\\
M4-S7&   4.30 &   0.02 &   34.3 &    1.4  &   -0.80 &  0.13 &   4805 &  100 &  12.26$_{-3.93} ^{+10.45}$ & 0.84 $_{-0.14}^{+0.09}$\\ \hline
\end{tabular} 
\endgroup
\end{table*}

\section{Distances and orbits} 
\label{Sect:DandO}
In this section we compute distances and orbits for the RAVE stars studied in this work. As a sanity check, we first compare distances estimates coming from five different methods, namely:
\begin{itemize}
\item scaling relation;
\item PARAM distances derived using UVES atmospheric parameters and the COR seismic values;
\item direct GAIA-DR2 parallax;
\item with the StarHorse pipeline \citep{starhorse}, using photometry and Gaia DR2 data, assuming a parallax zero-point correction of 0.52 mas (Zinn et al. 2018);
\item distances provided by \citet{BJ2018}.
\end{itemize}

Distances using scaling relations were derived using the expression of \citet{Miglio2013}, using the reddening as measured from \citet{Schlegel1998}:
\begin{equation}
\log{d}=1+2.5\log{\frac{T_{\rm eff}}{T_{\rm eff,\odot}}}+\log{\frac{\nu_{\rm max}}{\nu_{\rm max,\odot}}}-2\log{\frac{\Delta\nu}{\Delta\nu_\odot}}+0.2(m_{\rm bol}-M_{\rm bol, \odot})
\end{equation}
where $d$ is in parsec, $m_{\rm bol}$ is the apparent bolometric magnitude of the star, and $M_{\rm bol, \odot}$ the absolute solar bolometric magnitude. Bolometric corrections were adopted from \citet{BC2018}. Errors are calculated using propagation of uncertainty. Distances calculated using the different methods listed above are summarised in Table~\ref{Tab:seismgaia}. 

\begin{table}
\begingroup
\renewcommand{\arraystretch}{1.2} 
\centering
\caption{Reddening values for each stars as calculated from \cite{Schlegel1998}, PARAM and COR seismic values, StarHorse \cite{Queiroz2018} (spectroscopic atmospheric parameters and Gaia parallaxes) and \cite{Green2018}.}
\begin{tabular}{llllc}  
\hline \hline         
Star & Av$_{\rm Schl.1998}$ & Av$_{\rm PARAM}$          & Av$_{\rm StarHorse}$    & Av$_{\rm Green2018}$ \\ 
     &   [mag]              & [mag]                     &    [mag]                &   [mag]      \\ \hline
S1   & 0.079                & 0.506$_{-0.381} ^{+0.601}$& 0.494$_{-0.018} ^{+0.006}$& 0.093 $\pm$ 0.062\\
S2   & 0.133                & 0.715$_{-0.165} ^{+0.408}$& 0.551$_{-0.017} ^{+0.014}$& 0.155 $\pm$ 0.062\\
S3   & 0.123                & 0.170$_{-0.115} ^{+0.170}$& 0.099$_{-0.024} ^{+0.226}$& 0.155 $\pm$ 0.062\\
S4   & 0.174                & 0.378$_{-0.113} ^{+0.006}$& 0.558$_{-0.011} ^{+0.011}$& 0.248 $\pm$ 0.062\\
\hline
\end{tabular}
 \label{Tab:reddening}
 \endgroup
\end{table}

The different distances are in broad agreement. In particular, SH distances assuming a parallax zero point of $-$0.52 mas (Zinn et al. 2018) are in good agreement with those obtained from PARAM. In Table~\ref{Tab:reddening} the SH estimate extinctions for the four stars are given and compared with values from the literature.  In the rest of our analysis we will adopt the PARAM distances.

Orbit parameters were calculated using GALPY \citep{Bovy2015} \footnote{Code available at \url{http://github.com/jobovy/galpy}.}. We adopted a Galactic potential (MWpotential2014) and a solar radius of 8.3 kpc. We adopted PARAM distances and, when available, Gaia proper motions (see Tables~\ref{Tab:data} and \ref{Tab:seismgaia}). In the case of 201359581 (S1) we adopted PARAM distances and UCAC 5 \citep{UCAC5} proper motions, since the Gaia astrometric solution is not reliable. Errors on orbit parameters were calculated via Monte-Carlo approach, simulating 1,000 stars per object with velocity, distance and proper motions varying within errors. Results are summarised in Table~\ref{Tab:orbits}.
\begin{table*}
\begingroup
\renewcommand{\arraystretch}{1.2} 
\caption{Adopted proper motions for the orbit integration, plus orbit parameters of the stars in this work. Distance has been derived by PARAM, using BM seismic parameters (see Table~\ref{Tab:seismres}); radial velocity has been measured from ESO spectra via cross-correlation (see Table~\ref{Tab:abundances})and proper-motions were taken from GAIA-DR2 catalogue (UCAC-5 for S1, due to the flags in Gaia DR2 catalogue). Orbits have been integrated using Galpy v.1.4.0 using \it MWpotential2014 potential.}
\label{Tab:orbits}
\centering
\begin{tabular}{lrrrrrrrrr}
\hline \hline
ID    & PMRA                   & PMDE                 & U                                & V                             & W                          &  Rmin                & Rmax                   & ecc  & Zmax   \\ 
      & mas/yr                 & mas/yr               & km/s                             & km/s                          & km/s                       & kpc                  & kpc                    &      & kpc    \\ \hline
S1    & $-$51.2 $\sigma$= 0.9  & $-$4.7 $\sigma$= 1.0 & $-$307.9$_{-23.9} ^{+10.8}$    & $-$225.2$_{-13.7} ^{+7.9}$  & $-$34.4$_{-11.3} ^{+8.9}$   & 0.5$_{-0.5} ^{+1.2}$   &  24.2$_{-5.9} ^{-2.4}$     & 0.96$_{-0.03} ^{+0.02}$ &  9.2$_{-0.6} ^{+1.1}$  \\
S2    &  16.79 $\sigma$=0.49     & $-$4.50 $\sigma$=0.18  & $-$183.7$_{-25.0} ^{+12.9}$    & $-$155.1$_{-25.0} ^{+13.4}$ &   89.8$_{-19.4} ^{+13.5}$  & 1.6$_{-0.9} ^{+1.8}$   &  11.5$_{-1.2} ^{-1.5}$     & 0.75$_{-0.08} ^{+0.06}$ &  9.0$_{-1.8} ^{+2.1}$   \\
S3    &  $-$24.20 $\sigma$=0.06  & $-$48.96 $\sigma$=0.05 &    289.4$_{-21.1} ^{+11.7}$    & $-$240.0$_{-24.2} ^{+15.9}$ & $-$43.2$_{-3.6} ^{+11.0}$   & 0.1$_{-0.2} ^{+0.6}$   &  22.5$_{-6.8} ^{-1.2}$     & 0.98$_{-0.03} ^{+0.01}$ &  8.7$_{-1.4} ^{+1.1}$ \\
S4    &  32.43 $\sigma$=0.07     & 0.85 $\sigma$=0.07  &  $-$104.6$_{-6.9} ^{+6.6}$      &  $-$45.6$_{-3.8} ^{+4.9}$   & $-$21.1$_{-5.4} ^{+6.5}$    & 4.3$_{-1.1} ^{+0.6}$   &   10.03$_{-1.2} ^{-0.7}$     & 0.39$_{-0.03} ^{+0.05}$ &  0.6$_{-0.1} ^{+0.2}$   \\ \hline
\end{tabular}
\endgroup
\end{table*}

Three out of four stars are on very eccentric orbits, attaining large distances, typical of what is expected for halo stars. Figure~\ref{Fig:toomre} shows that 3 of the four studies stars occupied the halo locus in the Toomre diagram, whereas 206443679 (S4, our most metal poor star and the star with the less eccentric orbit) seems to be more consistent with a thick disk kinematics.

   \begin{figure}
   \centering
   \includegraphics[width=0.8\columnwidth]{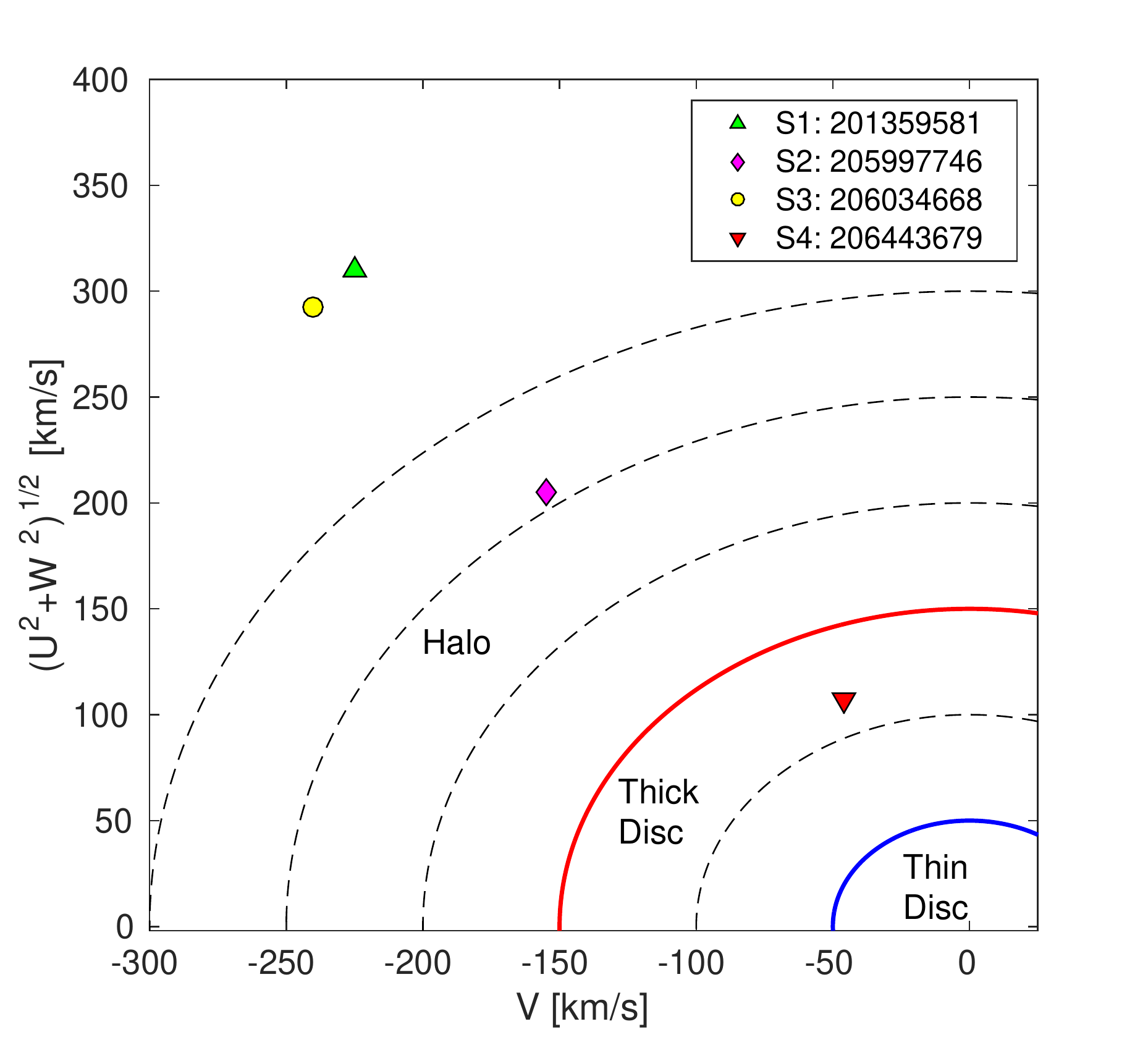}
   \includegraphics[width=0.8\columnwidth]{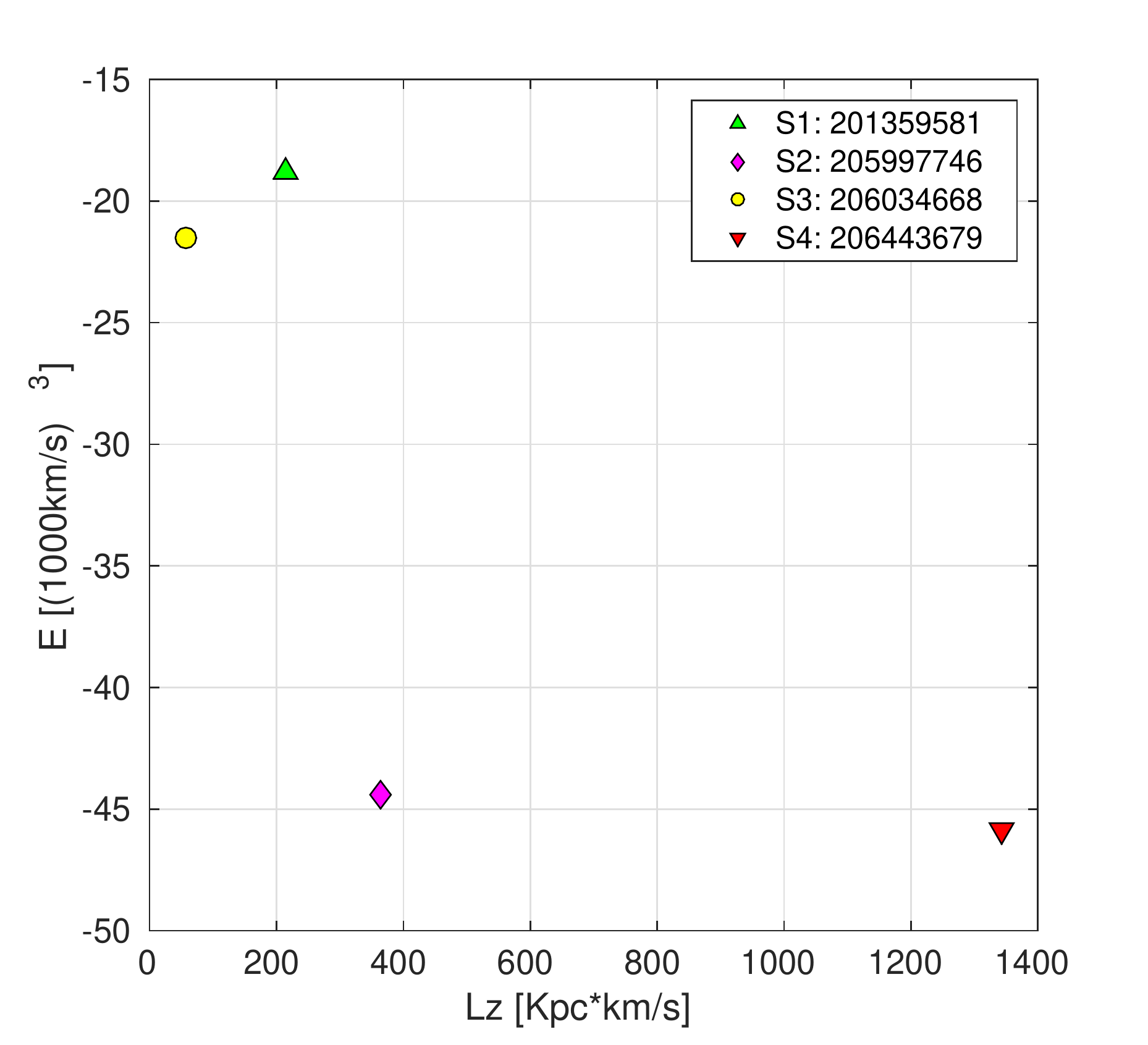}
   \caption{Top panel: Toomroe diagram of the RAVE stars of this paper. Indicative limits for the thin and thick disk are plotted. Bottom panel: Orbit energy vs Lz plot of the stars.}
              \label{Fig:toomre}%
    \end{figure}

\section{Summary of the properties of the four metal-poor RAVE stars}
In this section we give a brief summary of the main properties of each of the four RAVE stars, by combining all the information we obtained: chemistry, ages and masses, and kinematics.

\label{Sec:discussionindiv}

\subsection{201359581 (S1)}
This object is the only star of the sample where the temperature derived from the high-resolution spectrum is 380 K lower than the \Teff derived from the lower resolution RAVE spectrum and the \Teff derived from the IRFM. We already noticed in \citet{Valentini2017} that the IRFM tends to overestimate temperatures at \Teff $>$ 5000 K. This is probably due to the adoption of RAVE parameters as an input in the IRFM from RAVE-DR5. The miscalculated temperature from the RAVE spectrum led to a underestimated age for this star (see Appendix~\ref{RAVEages}). The \Teff derived from high-resolution spectroscopy brings the age back into agreement with the expectation of this very metal-poor star being old. This is the star with the lower value of \Dnu~ and \numax, and, looking at its position in the HR diagram, Fig.~\ref{Fig:tracks}, it is the only object that could be confused with a red-clump star, which would then contribute to more uncertain estimates of mass and radius, and therefore age (mostly due to mass loss). The PDF of the age has a complex profile, multi-peaked.
Among the four stars, this is the object with the largest [C/Fe] ratio (around 0.30 dex). The star has both a high Ba and Eu also a [Eu/Ba] ratio of 0.3 $\pm$0.11. The small variation (few km/s) in radial velocity and the big error ($>$5 km/s) associated to Gaia radial velocity suggest that this object can be a binary star.
Due to the flags in the astrometric solutions, the orbital parameters obtained for this star are larger, since we adopted the less precise proper motions from UCAC-5 catalogue \citep{UCAC5}. The star has an highly eccentric orbit and, looking at the Toomroe diagram in Fig.~\ref{Fig:toomre}, it can be classified as an Halo star.

\subsection{205997746 (S2)}
The star is not C enhanced, it is below the RGB bump, see Fig.~\ref{Fig:tracks}.
The star appears enhanced in Na: [Na/Fe]=$+$1.18 dex, but this result has to be taken carefully. 
This might be due to unresolved Na interstellar lines that hamper the abundance measurement. 
For this reason we are adopting this value as an upper limit.
The star is alpha enhanced, and it is rich in Eu ([Eu/Fe]=0.41) and it can be classified as an 
r-rich star ([Eu/Fe]$>$0.3). The star is the richest in Cu (and poor C) of our sample. 
The age of this star presents a multi-peaked PDF, as 
visible from Fig.~\ref{Fig:BM_vs_BMN} and Fig.~\ref{Fig:compAgeLim}.
Looking at the kinematics of the object, Fig.~\ref{Fig:toomre}, the star seems a typical Halo star.

\subsection{206034668 (S3)}
Looking at the HR diagram in Fig.~\ref{Fig:tracks}, the star is located below the bump. 
It is alpha-enhanced and it does not show C-enhancement. This star is the richest star in Ba of our sample, 
while [Eu/Fe] is almost solar (r-poor). The low C-abundance and the absence of v$_{\rm rad}$ variation that 
might indicate binarity, suggest that the star is not Ba-enriched via mass transfer from a more massive companion 
while in AGB phase. If we use the element ratios as a diagnostic we find [Eu/Ba]=$-$0.89$\pm$0.12 and [Sr/Ba]=$-0.89\pm0.15$. 
Following \cite[][Fig.~4]{Spite2018}, these ratios put the star outside the correlation of [Eu/Ba] and [Ba/Fe], suggesting 
an origin from an environment with different chemical history than the Galactic Halo.
When looking at mass and age of 201034668, PARAM provides different results depending on the seismic pipeline adopted. 
COR seismic values provided a double-peaked age PDF, with no probability that the star is younger than 4 Gyr, when GRD seismic 
values lead to older age. In all the cases the age PDF extends beyond 30 Gyrs (or truncated when the age prior is adopted, 
as visible from Fig.~\ref{Fig:compAgeLim}. 
The star seems to have a slightly retrograde orbit: in the Toomroe diagram the star is beyond the $-220$ km/s (the slightly retrograde 
orbit is maintained also when integrating the orbit using Gaia-DR2 distances). This star 
has an angular velocity  of v$\phi_{mean}$=$-$0.133 km/s and it is on a highly energetic orbit, looking at the lower panel 
of Fig.~\ref{Fig:toomre}. 
The Ba and Eu enrichment, combined with the retrograde orbit, suggests that this star might be accreted from a system with 
larger Ba enrichment, such as a dSph galaxy \citep{Spite2018}. 

\subsection{206443679 (S4)}
The star is well located below the RGB bump.
The alpha-enrichment, the high Eu-content ([Eu/Fe]=0.79dex), and the low C-content, suggest that the star is chemically a typical r-rich Halo star. The star has also an high Sr and Ba-content ([Ba/Fe]=0.83 dex;[Sr/Fe]=0.69 dex), giving [Sr/Ba]=$-$0.14. Following \cite[][Fig.~4]{Spite2018}, this puts the star very close to the pure r-process production limit.
The star has an orbit typical of a thick disk star, however, its metallicity of [Fe/H] = -2.2 dex is indicative of a halo/accreted origin. This star could have acquired the presently observed orbit in two ways:
\begin{enumerate}
\item Keeping in mind its age of 9-10 Gyr, it could have belonged to Milky Way’s last massive merger. It can be seen in Fig.1 of Helmi et al. (2018) that this region of the Toomre diagram is degenerate with respect to accreted and in-situ born population. This requires an in-plane accretion, which can result from massive mergers being dragged into the disk mid-plane by dynamical friction \citep{Read2009}.  
\item The inner halo has long been known to acquire angular momentum from the bar, causing it to slows down, as seen in N-body simulations bar as (e.g. \citealp{Athanassoula2003, Minchev2012}). With a guiding radius of ~7 kpc, this star may have therefore gained rotational support from the bar.  
\end{enumerate}

\section{Conclusions}
\label{Sect:conclusions}

As part of a pilot program aimed at obtaining precise stellar parameters and ages for very metal-poor stars with available seismic information, we here determined mass and ages for a sample of 4 RAVE metal-poor stars. We also characterized the stars by combining the information on age, with their chemical profile (form high-resolution UVES spectra, covering different production channels) and their kinematics.
Our analysis took advantage of the seismic information derived from K2 light curves (Campaigns 1 and 3): asteroseismology was first involved in the spectroscopic analysis and then in the mass and age determination using a Bayesian approach. 
We provided a full analysis (stellar parameters, chemistry and ages) using both intermediate-resolution spectra (RAVE, R= 7,500) and high-resolution spectra (ESO-UVES, R= 110,000). 
We found abundances and atmospheric parameters derived from the high-resolution spectra to be in agreement with the atmospheric parameters derived form RAVE spectra once our strategy of making use of the seismic gravities and iterating on a more consistent (\logg, \teff) pair, is adopted, as described in \citet{Valentini2017}. 

In addition we provide a comparison of \Logg derived using three different methods: a) from the classical spectroscopic analysis, b) from Gaia DR2 parallaxes (Eq.~\ref{Eq:parlogg}), c) from asteroseismology (Eq.~\ref{Eq:sismologg}). The three estimated values are in agreement within errors and seismic \Logg demonstrated to be reliable even at low metallicities, with the advantage of providing the most precise measurement. At low metallicities the classical \Logg derived via ionisation equilibrium is affected by NLTE effects, that may hamper the correct estimate of gravity and temperature. The \logg$_\varpi$, even if it has a large uncertainty due to the mass assumption, can be used as a good prior for spectroscopic analysis of red giant spectra, in particular of spectra with known \teff-\Logg degeneracies (as in RAVE) when no seismic information is available. 

The more precise and self-consistent stellar parameters obtained for the four RAVE stars, when combined with \Dnu~ and \numax~ estimated from different seismic pipelines, deliver masses and ages with 9\% and 30-35\% uncertainties, respectively. Ages for field red giants of this precision opens new perspectives to the field of Galactic Archaeology \citep[see also]{Miglio2017}. 
Along this work we also investigated the impact of different assumptions on the above uncertainties. The main conclusions can be summarised as follows:
\begin{itemize}
\item Impact of spectral resolution/short-wavelength interval: masses and ages were obtained from RAVE and UVES spectra using the same strategy of iterating on the best (log g , Teff) pair using as prior the seismic gravity and the IRFM temperatures. In the case of the RAVE spectra the known degeneracies lead to large uncertainties in mass and age. In one case, when the IRFM \Teff was inaccurate by $\sim$ 250 K (i.e. outside the flexibility range in temperature during the iteration) an erroneous age determination occurs. However, in this case, the posteriors of temperature, mass, and age are in tension with those of \Dnu and \numax. This already tends to indicate an erroneous determination on one of the input parameters, and thus potentially leading to an erroneous age determination (this case is well illustrated by that specific example). 
\item Impact of the different seismic pipelines: the adoption of different seismic pipelines has made clear the important impact the uncertainties on \Dnu and \Numax estimates can have on the resulting masses and ages. However, also in this case, it is possible to select those seismic estimates that seem in better agreement with the quality of the light curves available, making sure that only the best seismic parameters are used. In this work we favoured the seismic method providing the lowest spread when compared to other methods \citep[][Fig.~10]{Pinsonneault2018}.
\item The impact of surface temperature scale:  a shift of $-$100K in \Teff leads to a mass underestimation of $\sim$ 10\% and, as consequence, a stellar age that is older by $\sim$30\% (if temperatures are overestimate the effect works in the opposite direction.);
\item The impact of [$\alpha$/Fe] ratios: In this case the impact is less important than the ones discussed above (being only of a few \% in age). In the case of the RAVE spectra, where the [$\alpha$/Fe] has larger uncertainties, we have computed ages and masses for different  [$\alpha$/Fe] ratios and the effects were minor. 
\item The impact of mass-loss: as pointed out in Anders et al. (2017) and Casagrande et al. (2016), the impact of mass-loss becomes important in the RC phase. Our stars are compatible with being RGB where the mass-loss impact is expected to be minor (as also shown by the computation made in the present work).
\item The impact of an accretion event: the seismic age measurement relies on the fact that the age of a red giant is proportional to the time spent on the MS, and therefore its mass. Any mass accretion event hampers this assumption (rejuvenating the star). Radial velocity variations (due to binarity) or chemical hints of mass transfer (mostly C or s-process elements contribution due to AGB-mass transfer) must raise a flag regarding the accuracy of the ages measured with asteroseismology. For three stars of our sample we have not find any clear sign of radial velocity variability, nor any clear chemical signature of mass transfer from a companion, and therefore we consider our ages reliable.
\end{itemize}

This pilot project shows that it is possible to use asteroseismology for determining precise and consistent masses and ages of metal-poor field giants. Together with nucleo-cosmo-chronometry, seismology provide the only way to estimate ages of distant field stars. However, this important new tool needs key steps to be followed, which are {\it i}) a consistent spectroscopic analysis which delivers not only detailed abundances, but also a consistent (\logg, Teff) pair; {\it ii}) a careful and critical use of the seismic inputs, and {\it iii}) an analysis of the posterior distributions of all output parameters to look for tensions with the seismic input which might be indicative of erroneous parameter estimates. The use of seismic \Logg and a temperature prior in an iterative way (see Valentini et al. and references therein), is thus a critical step in the analysis.
This important step assures that the atmospheric parameters used for deriving mass and age with asteroseismology is consistent with the seismic inputs used in the code, also offering a new way to provide more reliable surface temperatures.

In the near future the impact of the Gaia data should become important thanks to a better understanding of the parallax offsets and also in terms of narrowing the current posterior age distributions \citep[see][discussion]{Rodrigues2017}. For now, Gaia DR2 data are already useful to better define the orbits of the studied stars. 

Our strategy will enable a more serious program towards determining ages for giant halo field stars, that is complementary to nucleo-cosmo-chronometry, but with two advantages which are: it applies to all stars, and not necessarily only to those strongly r-process enhanced, and it provides ages with smaller uncertainties. Detailed abundance measurements are also necessary to gauge possible effects of mass-accretion which would systematically shift the seismic ages. Finally, the results of this pilot program pave the path for a more extensive study of metal poor stars with asteroseismology, delivering samples with age estimates to a  $\sim$30\% precision, hence superior to all what is currently available for field metal-poor distant stars in terms of age determinations. It seems not unrealistic to imagine that in the near future we will be able to add the age dimension in the chemical diagrams of the metal poor universe (e.g. Cescutti \& Chiappini 2014, Sakari et al. 2018, Spite et al. 2018), thus contributing enormously to our understanding of the first phases of the galaxy assembly and early nucleosynthesis.

\begin{acknowledgements}
MV and CC acknowledge the DFG project number 283705981: ``Analysing the chemical fingerprints left by the first stars: chemical abundances in the oldest stars''. DB is supported in the form of work contract FCT/MCTES through national funds and by FEDER through COMPETE2020 in connection to these grants: UID/FIS/04434/2019; PTDC/FIS-AST/30389/2017 \& POCI-01-0145-FEDER-030389. AM, WJC, GRD, and YPE acknowledge the support of the UK Science and Technology Facilities Council (STFC). TSR acknowledges  financial support from Premiale 2015 MITiC (PI B. Garilli). Authors MV, CC, DB acknowledge support from the “ChETEC” COST Action (CA16117), supported by COST (European Cooperation in Science and Technology). SM acknowledges support from NASA grant NNX15AF13G, NSF grant AST-1411685, and the Ramon y Cajal fellowship number RYC-2015-17697. The authors acknowledge the International Space Science Institute (ISSI-Bern), that hosted the first AsteroSTEP meetings (\url{https://www.asterostep.eu}). This work has made use of the VALD database, operated at Uppsala University, the Institute of Astronomy RAS in Moscow, and the University of Vienna. Based on data obtained from ESO-UVES instrument under proposal ID: 099.D-0913(A). Funding for RAVE has been provided by: the Australian Astronomical Observatory; the Leibniz-Institut fuer Astrophysik Potsdam (AIP); the Australian National University; the Australian Research Council; the French National Research Agency; the German Research Foundation (SPP 1177 and SFB 881); the European Research Council (ERC-StG 240271 Galactica); the Istituto Nazionale di Astrofisica at Padova; The Johns Hopkins University; the National Science Foundation of the USA (AST-0908326); the W. M. Keck foundation; the Macquarie University; the Netherlands Research School for Astronomy; the Natural Sciences and Engineering Research Council of Canada; the Slovenian Research Agency; the Swiss National Science Foundation; the Science \& Technology Facilities Council of the UK; Opticon; Strasbourg Observatory; and the Universities of Groningen, Heidelberg and Sydney. The RAVE web site is at: \url{https://www.rave-survey.org}. This work has made use of data from the European Space Agency (ESA) mission {\it Gaia} (\url{https://www.cosmos.esa.int/gaia}), processed by the {\it Gaia} Data Processing and Analysis Consortium (DPAC, \url{https://www.cosmos.esa.int/web/gaia/dpac/consortium}). Funding for the DPAC has been provided by national institutions, in particular the institutions participating in the {\it Gaia} Multilateral Agreement. We finally acknowledge the anonymous referee for the useful remarks.

\end{acknowledgements}

\begin{appendix}
\section{The selection of \Dnu and \Numax}
\label{SeismoChoice}
For the four RAVE metal-poor stars analysed in the present work we obtained \Dnu and \Numax from four different pipelines. We decided to select the best \Dnu and \Numax pair by looking at the performances of the four pipelines for each object. When looking at the power spectrum, see Fig.~\ref{Fig:numax}, it is visible that the uncertainty on the A2Z results is clearly too large in at least two instances. This is probably connected with the method and it's sensitivity to poorly sampled data. For this reason we will decide we do not favour the A2Z results for the \numax.

For better understanding the \Dnu results, SNR spectra have been created (Fig.~\ref{Fig:SNR}), then SNR spectra have been analysed as a function of frequency mod \Dnu divided by \Dnu (one realization per each pipeline). The same analysis has been performed using \Dnu$+$eDnu. If the uncertainty is sensible (i.e., not too large) we might expect to still see repeated structure. If e\Dnu is too large the repeated structure goes away.

This check led to the following conclusions regarding \dnu:
\begin{itemize}
\item 201359581 (S1), nothing is clearly visible in both SNR realisations.

\item 205997746 (S2) has a nice l=0,2 pair with all pipelines.

\item 206034668 (S3) has no result from A2Z, but the other three pipeline all find a result even if the epsilon value is not agreed on. Notice YE and GRD agree on epsilon. The sharpness of the peaks in this star seems to be better for GRD and YE rather than BM.

\item 206443679 (S4) is easy to see the l=0, 2 and plenty of other repeated structure. Every pipeline agrees for this star.
\end{itemize}

This is probably a result of the different method used and the degree to which the pipelines are set-up to be conservative. With only 4 stars we do not have the benefit of a large sample to cope with, having uncertainties that are too large. From the tests above we concluded that BM and GRD have the lowest and probably the most realistic uncertainties for these 4 stars (this conclusion does not necessarily hold for other stars). We therefore move forward with the analysis by using only the GRD and BM results. For the future works we will keep considering results from different pipelines, performing this analysis for every target.  

\begin{table}
\caption{\Dnu and \Numax as measured by the four different pipelines.}
\label{Tab:allseismo}
\centering
\begin{tabular}{lrrrrrrrrrrrrrrrr}
\hline \hline
\multicolumn{5}{l}{COR} \\ \hline
ID&\Numax   &e\Numax   &\Dnu      &e\Dnu     \\
  &[$\mu$Hz]&[$\mu$Hz] &[$\mu$Hz] &[$\mu$Hz] \\ \hline 
S1&20.2     &0.3       &2.79      &0.06      \\
S2&51.2     &1.1       &5.76      &0.06      \\
S3&41.8     &2.2       &5.26      &0.10      \\
S4&190.0    &8.0       &16.05     &0.06      \\ \hline
\hline
\multicolumn{5}{l}{GRD} \\ \hline
ID&\Numax    &e\Numax   &\Dnu      &e\Dnu    \\
  &[$\mu$Hz] &[$\mu$Hz] &[$\mu$Hz] &[$\mu$Hz]\\ \hline
S1&19.7      &0.8       &2.75      &0.16     \\
S2&50.0      &0.8       &5.80      &0.09     \\
S3&42.0      &1.3       &5.13      &0.10     \\
S4&188.5     &2.3       &16.15     &0.08     \\ \hline
\hline
\multicolumn{5}{l}{YE} \\ \hline
ID&\Numax    &e\Numax   &\Dnu      &e\Dnu    \\
  &[$\mu$Hz] &[$\mu$Hz] &[$\mu$Hz] &[$\mu$Hz]\\ \hline
S1&20.2      &0.5       &2.75      &0.07     \\
S2*&50.5      &0.9       &5.72      &0.07     \\
S3&43.8      &1.1       &5.06      &0.21     \\
S4&188.2     &2.2       &16.16     &0.11     \\ \hline
\hline
\multicolumn{5}{l}{A2Z} \\ \hline
ID&\Numax    &e\Numax   &\Dnu      &e\Dnu    \\
  &[$\mu$Hz] &[$\mu$Hz] &[$\mu$Hz] &[$\mu$Hz]\\ \hline
S1&20.8      &1.01      &--        &--       \\
S2&51.1      &2.46      &5.69      &0.15     \\
S3&42.3      &4.8       &--        &--       \\
S4&189.4     &24.23     &16.07     &0.05     \\ \hline
\end{tabular}
\begin{tablenotes}
      \small
      \item (*) For S2 the YE pipeline found that the \Dnu value is sensitive to the range in the spectrum used.  A central value is sprovided.  
    \end{tablenotes}
\end{table}

\begin{figure}
\centering
\includegraphics[width=1.\columnwidth]{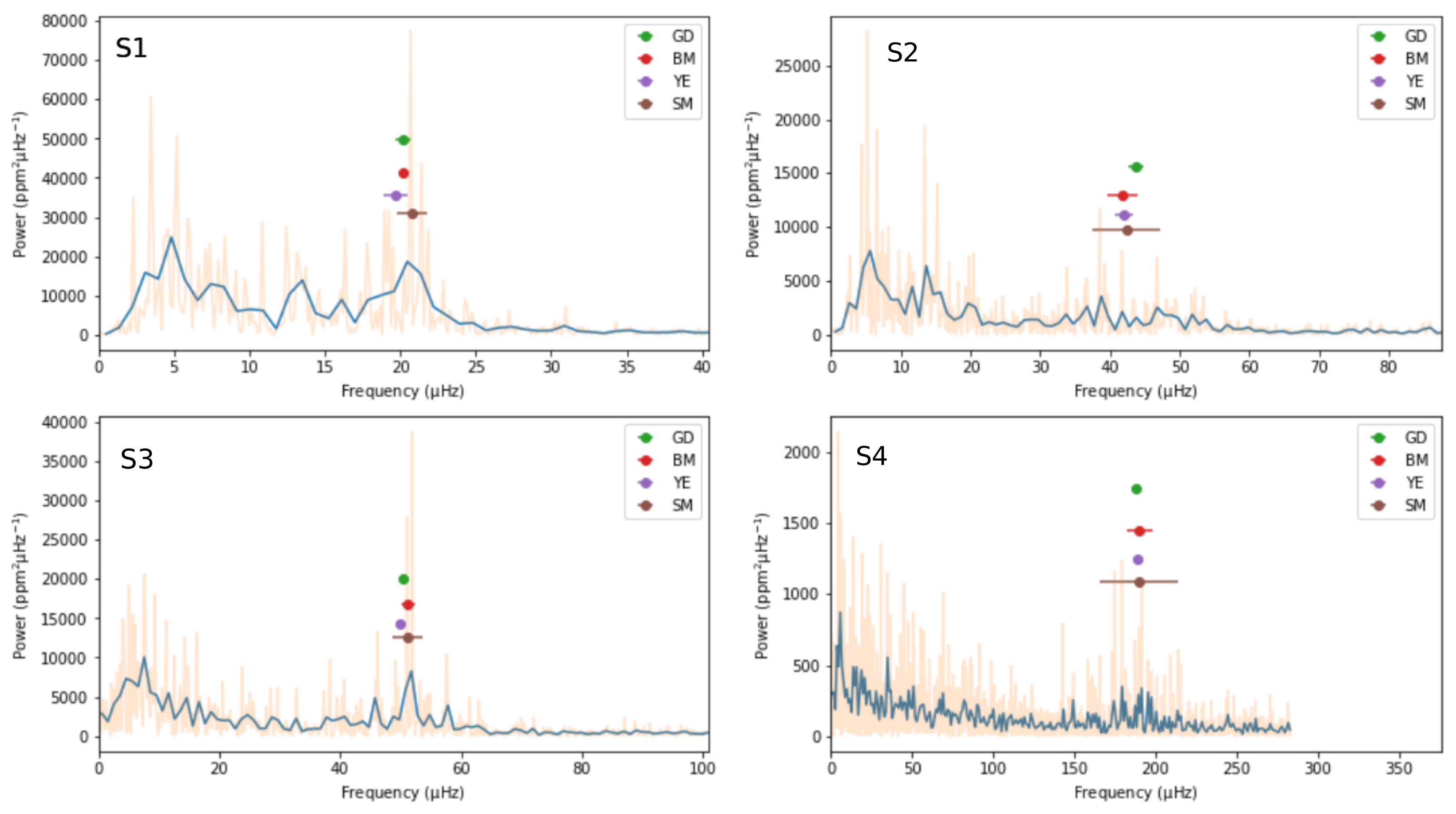}
\caption{Normalized power spectrum of the RAVE stars (original: orange line, smoothed: blue). \Numax values from the different pipelines are plotted: GRD (green), COR (red), YE (purple), and A2Z (brown).}
\label{Fig:numax}
\end{figure}
\begin{figure}
\centering
\includegraphics[width=1.\columnwidth]{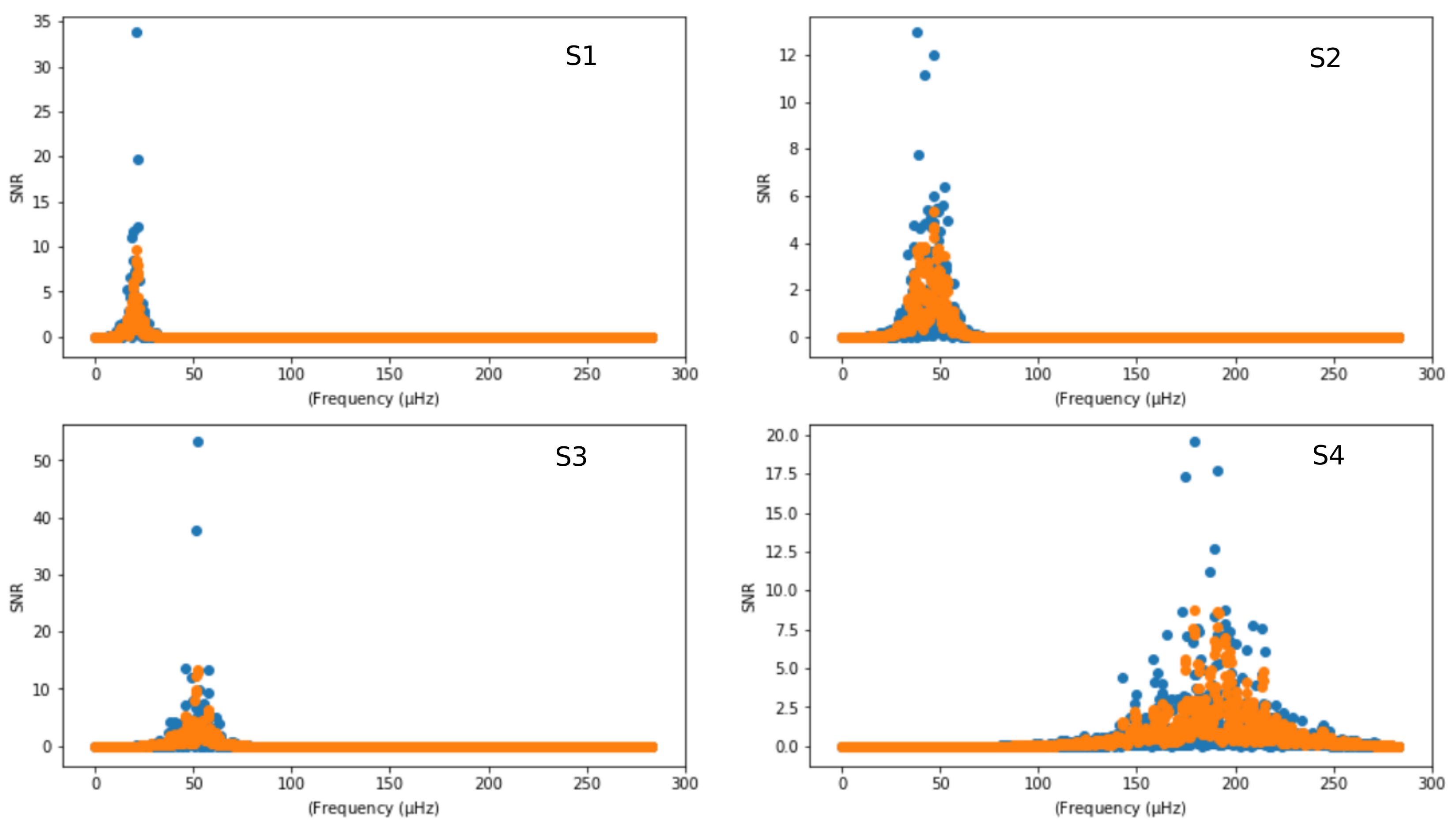}
\caption{SNR spectra of the RAVE metal-poor stars of this work.}
\label{Fig:SNR}
\end{figure}

\section{Analysis of UVES spectra}
\label{UVESappendix}

\begin{table*}
\caption{Chemical abundances derived for the metal-poor stars presented in this work. Values were derived from UVES spectra via equivalent width measurement (ew) or line fitting (f) using the atmospheric parameters derived using the seismic \logg. }
\label{Tab:abundances}
\centering          
\begin{tabular}{ll|ccc|ccc|ccc|ccc|c}     
\hline  \hline 
      &  &   \multicolumn{3}{|c|}{201359581 - S1} &  \multicolumn{3}{|c|}{205997746 - S2} &  \multicolumn{3}{|c|}{206034668 - S3} &  \multicolumn{3}{|c}{206443679 - S4}&  \\ \hline
Species & At. N. &  Nlin&   Abd  &  eA  &  Nlin  & Abd. & eA &  Nlin  & Abd. & eA  &  Nlin  & Abd. & eA & Met.\\
C    & 6.0    &      & 6.92   & 0.15  &   & 6.94 & 0.10 &   & 7.08 & 0.11 &   & 6.48 & 0.09 & f \\
Na I &11.0    & 2    & 4.71   & 0.06  & 2 & 6.07 & 0.12 & 2 & 4.96 & 0.05 & 2 & 4.68 & 0.08 & ew \\
Mg I &12.0    & 3    & 6.24   & 0.11  & 3 & 6.92 & 0.05 & 3 & 6.34 & 0.15 & 3 & 6.36 & 0.10 & ew \\
Si I &14.0    & 2    & 6.45   & 0.07  & 2 & 6.81 & 0.03 & 2 & 6.60 & 0.10 & 2 & 6.36 & 0.10 & ew \\
Ca I &20.0    &  7   & 5.01   & 0.05  & 7 & 5.45 & 0.10 & 7 & 5.05 & 0.13 & 7 & 4.95 & 0.13 & ew \\
Sc II&21.1    &  3   & 1.66   & 0.11  & 3 & 1.84 & 0.14 & 3 & 1.74 & 0.11 & 3 & 1.53 & 0.14 & ew \\
Ti I &22.0    &  11  & 3.40   & 0.08  & 11& 3.91 & 0.15 & 9 & 3.59 & 0.10 & 11& 3.58 & 0.10 & ew \\
Ti II&22.1    &  17  & 3.56   & 0.10  & 17& 3.95 & 0.11 & 16& 3.68 & 0.12 & 17& 3.60 & 0.09 & ew \\ 
Cr I &24.0    &   9  & 3.84   & 0.22  & 9 & 4.38 & 0.11 & 7 & 3.98 & 0.09 & 9 & 3.87 & 0.09 & ew \\
Cr II&24.1    &  --  & --     & --    & 2 & 4.70 & 0.08 & 2 & 4.04 & 0.08 & 2 & 4.18 & 0.07 & ew \\ 
Mn I &25.0    &  1   & 3.69   & 0.08  & 1 & 4.32 & 0.10 & 1 & 3.82 & 0.09 & 1 & 3.48 & 0.08 & ew \\
Fe I &26.0    & 69   & 5.59   & 0.11  & 68& 6.16 & 0.10 & 67& 6.05 & 0.12 & 63& 5.51 & 0.10 & ew  \\
Fe II&26.1    & 10   & 5.63   & 0.10  & 9 & 6.18 & 0.09 & 7 & 6.11 & 0.10 & 7 & 5.61 & 0.12 & ew \\
Ni I &28.0    & 3    & 4.57   & 0.10  & 3 & 5.08 & 0.09 & 3 & 4.67 & 0.11 & 3 & 4.49 & 0.07 & ew \\  
Cu I &29.0    & 1    & 2.19   & 0.07  & 1 & 3.06 & 0.10 & 1 & 2.49 & 0.08 & 1 & 2.03 & 0.10 & ew \\                           
Zn I &30.0    & 2    & 3.05   & 0.11  & 2 & 3.51 & 0.11 & 2 & 3.42 & 0.14 & 2 & 3.00 & 0.11 & ew \\                          
Sr I &38.0    &  1  & 1.16  & 0.08  & 1 &1.47  & 0.09 & 1 & 1.37& 0.11& 1 &1.64 & 0.11& f \\                            
Ba II&56.1    &  2  & 0.87 & 0.08  &3 & 1.18 & 0.09& 2 & 1.57 & 0.10 & 2 & 1.05& 0.13 & f \\  
Eu II&63.1    &  1  & $-$0.49  & 0.07  & 1 & $-$0.38 & 0.08 & 1 & $-$0.98 & 0.08 & 1 & $-$0.66 & 0.08& f \\
Gd II&64.1    &  1  & $-$0.69  &0.07   & 1 &$-$0.58 & 0.08& 1&$-$0.03 &0.10 & -- & --& --& f \\ \hline
\end{tabular}
\end{table*}

\section{Tests on alpha-enhancement and \Teff shifts}
\label{AlphaJust}

In the present work we use PARAM which uses a set of MESA models, not $\alpha$-enhanced. The effect of the $\alpha$-enhancement is taking into account by adopting the Salaris formula in Equation~\ref{Eq:salaris}. We tested this assumption using PARSEC models, for which alpha-enhanced computations are available.
In Fig.~\ref{Fig:alphatest} we compare two sets of PARSEC tracks, covering MS to RGB tip phases. One set (plotted in red) is a set of tracks for 0.7, 1.0 and 1.3 \Msun\ at \FeH $=-2.16$ dex and [$\rm \alpha/Fe$]$=0.4$ dex, a set of tracks at the corresponding metallicity \FeH$=-1.86$ dex (following Equation~\ref{Eq:salaris}). The maximum deviation between the two set of tracks reaches the maximum in the RGB phase. Since the difference is negligible respect to the typical errors we have on age, we adopted the \cite{Salaris1993} correction in our computations.

\begin{figure}
\centering
\includegraphics[width=1.\columnwidth]{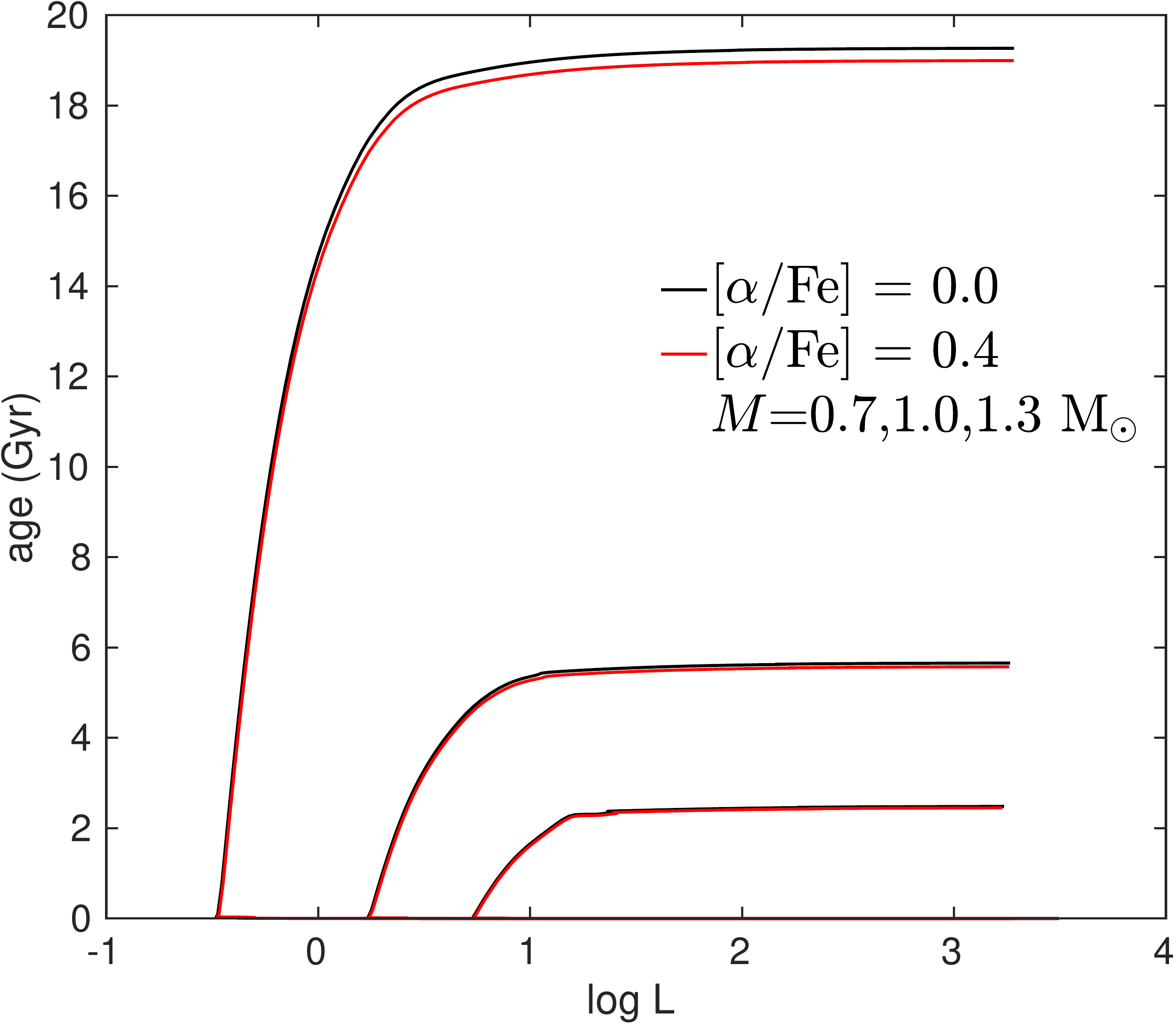}
\caption{PARSEC tracks at \FeH=$-$2.16 dex and  [${\rm\alpha/Fe}$]=0.4 dex (in red) and PARSEC tracks at the corresponding metallicity \FeH=$-$1.86, computed using Salaris formula (in black). Only MS-RGB tip is plotted and 3 masses are considered: 0.7, 1.0, and 1.3 \Msun.}
\label{Fig:alphatest}
\end{figure}

\begin{figure}
\centering
\includegraphics[width=1.\columnwidth]{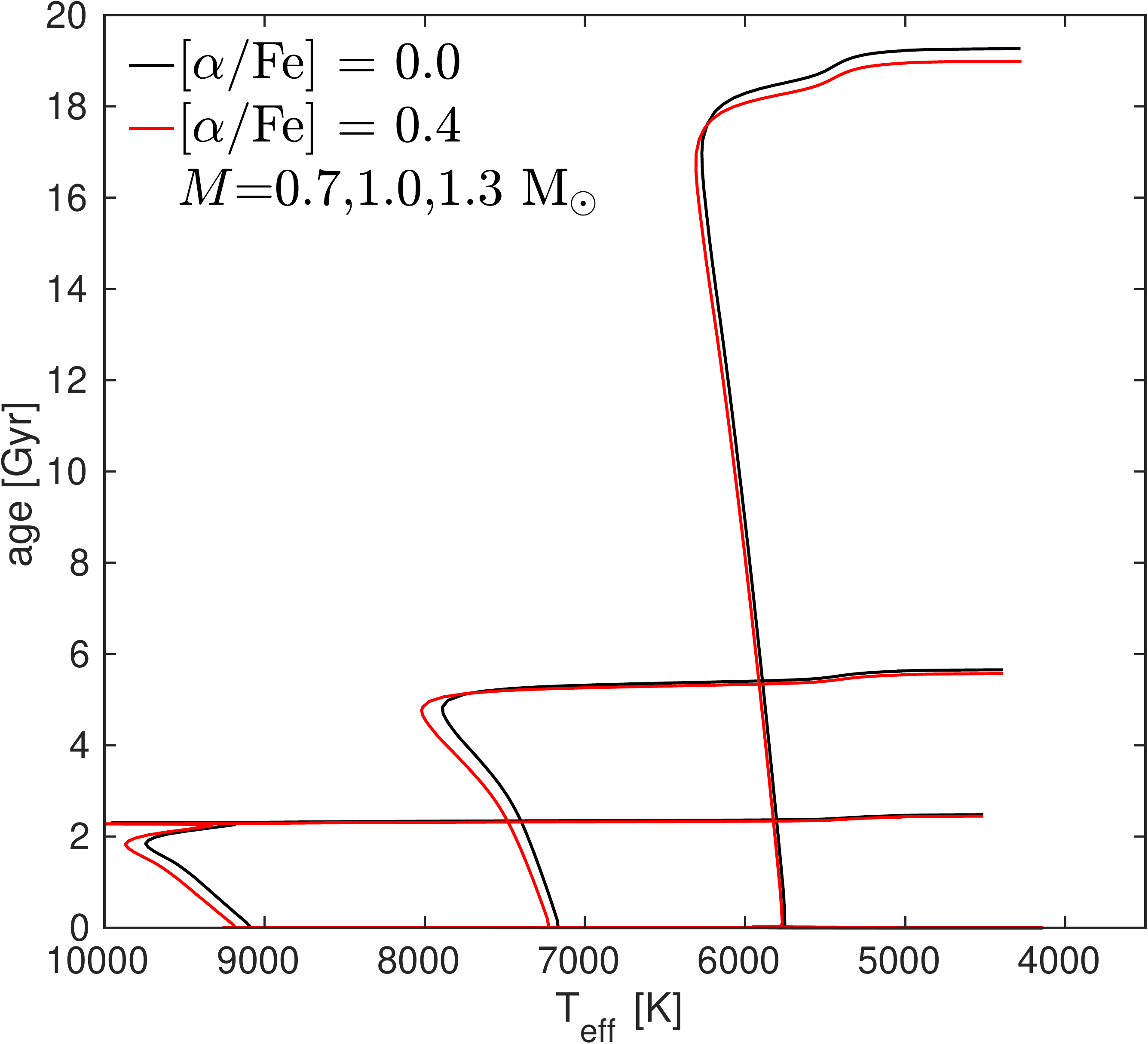}
\caption{Same PARSEC tracks of Fig.~\ref{Fig:alphatest} but with temperature on the abscissa.}
\label{Fig:temptest}
\end{figure}

\section{Masses and ages using RAVE atmospheric parameters and metallicities}
\label{RAVEages}

We derived ages and masses for the four RAVE metal poor stars using the atmospheric parameters derived from RAVE spectra using the seismic \logg. RAVE spectra cover a small spectral range (8420-8780\AA) at intermediate resolution (R=7,500), element abundances may suffer of offsets and inaccuracies. For this reason we computed ages and masses for five different $\alpha$ enhancements. Two different mass-loss approximations ($\eta$=0.2 and 0.4) have been considered and we adopted COR and GRD seismic parameters. Masses and ages derived using parameters measured from RAVE spectra are shown in Fig.~\ref{Fig:agesBenoit} (COR) and \ref{Fig:agesGuy} (GRD). The impact of temperature shift on this set of data has been tested by varying the \Teff of $\pm$100 K (see Fig.~\ref{Fig:Benoitalltests}) using COR seismic parameters.

From Fig.~\ref{Fig:agesBenoit}, \ref{Fig:agesGuy} and \ref{Fig:Benoitalltests} it is possible to see the effect of the different $\alpha$-enhancement and mass-loss assumptions, and the effects of shifts in \Teff.

\begin{figure*}
\centering
\includegraphics[width=1.7\columnwidth]{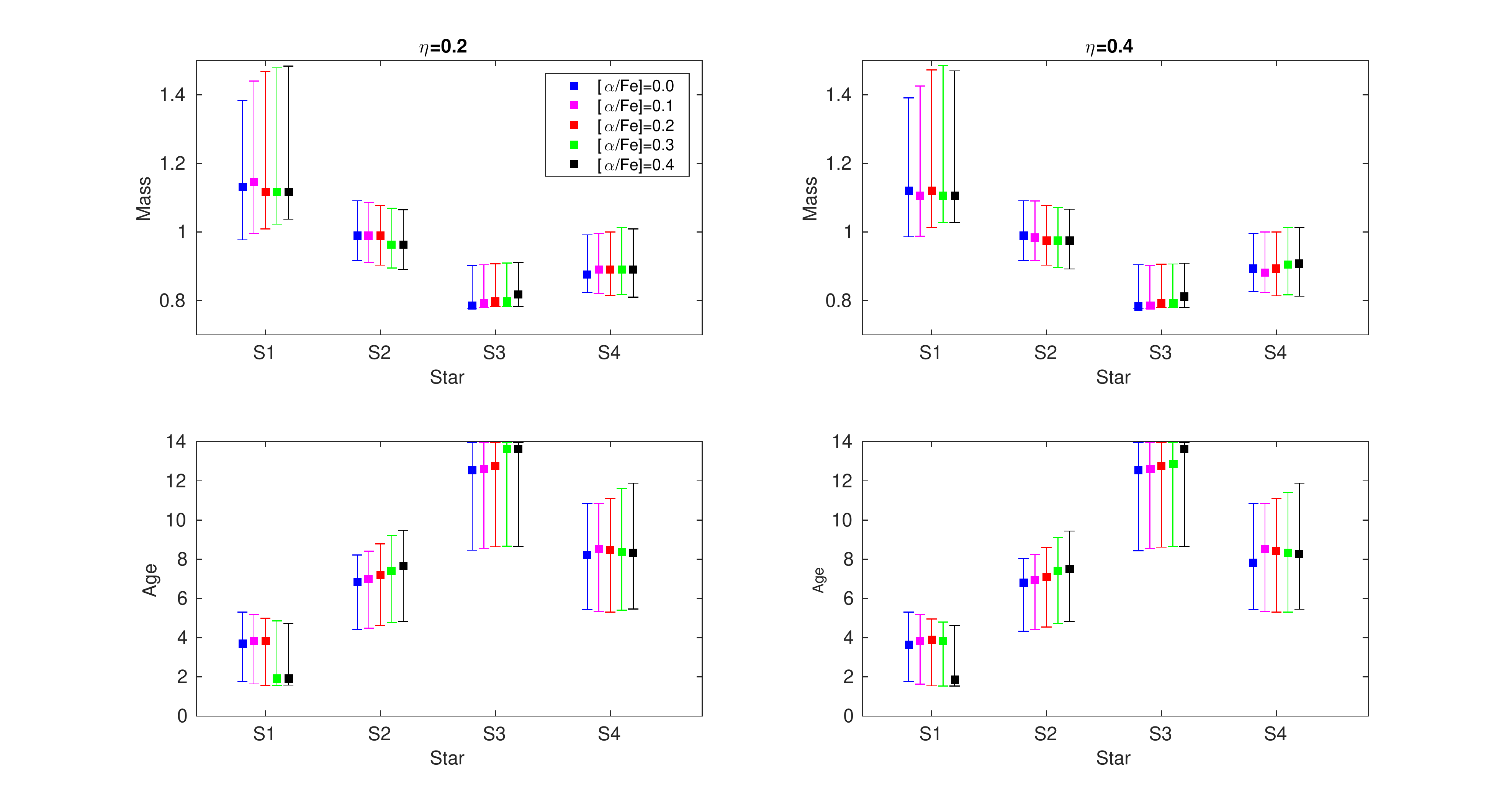}
\caption{Mass and ages of the five stars,determined using different [$\alpha$/Fe] (0.0, 0.1, 0.2, 0.3, 0.4 dex respectively) and two different $\eta$ parameters (0.2 and 0.4) for the mass loss. COR seismic parameters and spectroscopic parameters derived from RAVE spectra and asteroseismology.}
              \label{Fig:agesBenoit}
\end{figure*}


\begin{figure*}
\centering
\includegraphics[width=1.7\columnwidth]{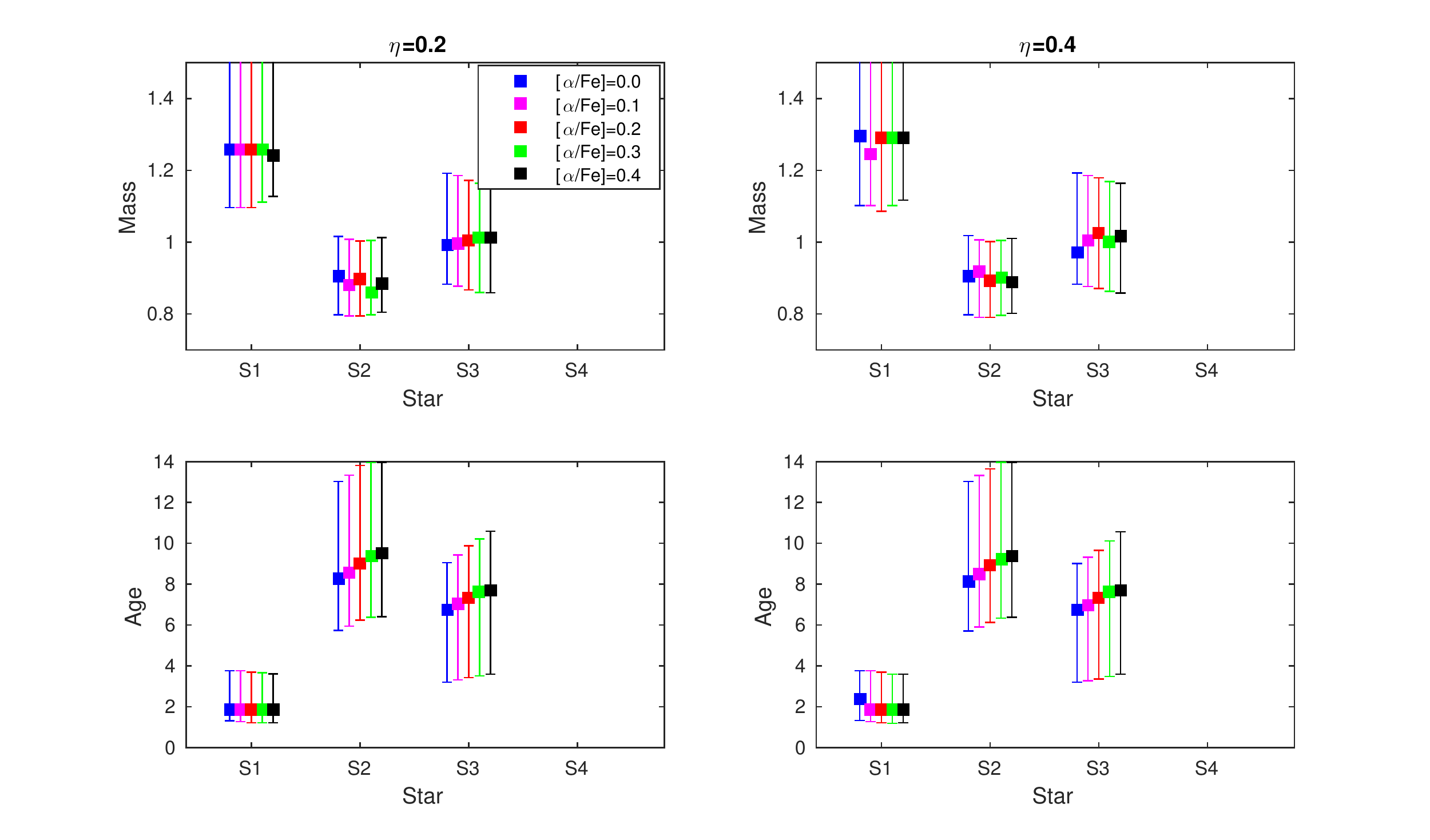}
\caption{Mass and ages of the five stars,determined using different [$\alpha$/Fe] (0.0, 0.1, 0.2, 0.3, 0.4 dex respectively) and two different $\eta$ parameters (0.2 and 0.4) for the mass loss. GRD seismic parameters and spectroscopic parameters derived from RAVE spectra and asteroseismology.}
              \label{Fig:agesGuy}
\end{figure*}

   \begin{figure*}
   \centering
   \includegraphics[width=1.89\columnwidth]{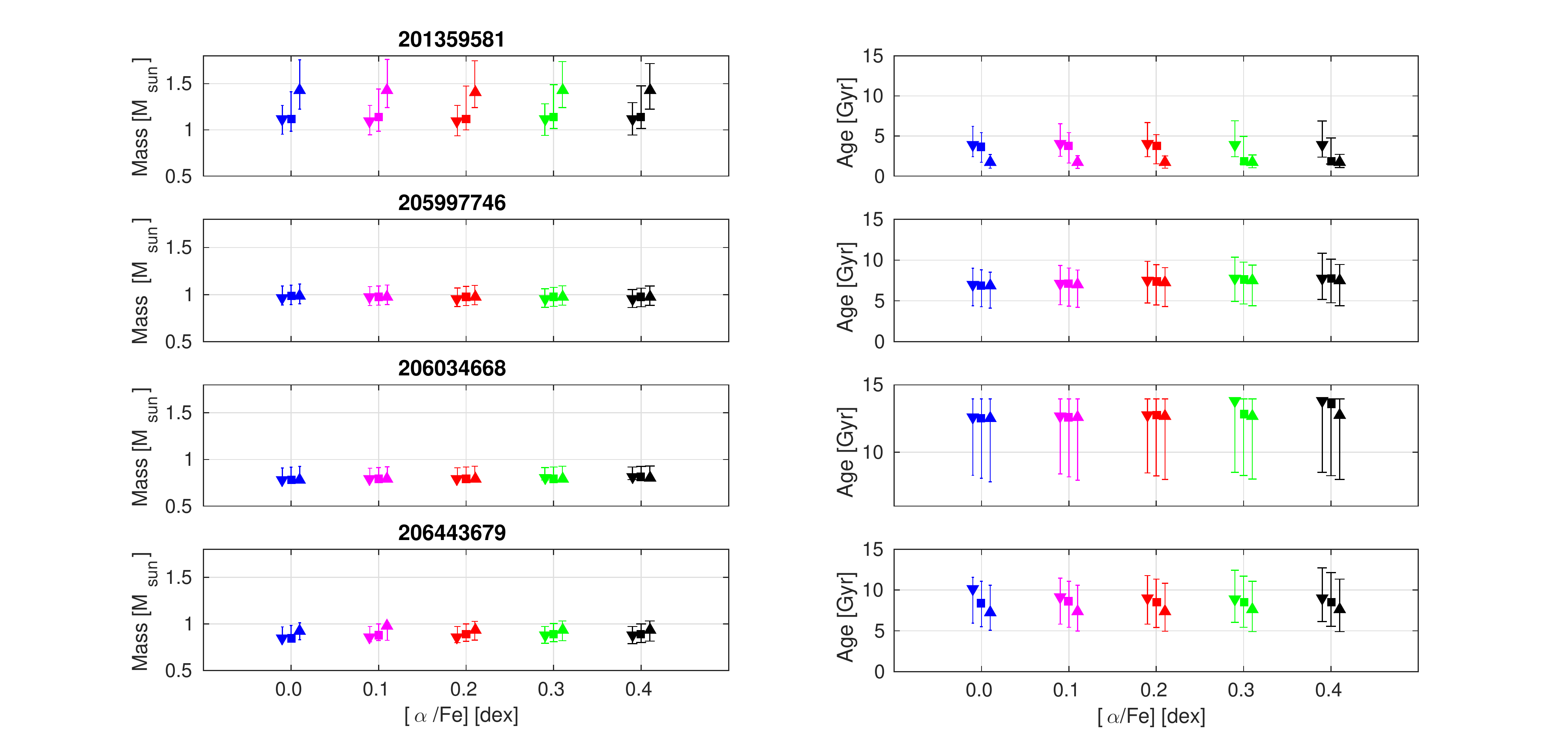}
   \caption{Mass and ages of the 5 stars,determined using different [$\alpha$/Fe] (0.0, 0.1, 0.2, 0.3, 0.4 dex respectively in blue, magenta, red, green and black) and varying the \Teff  of $\pm$100 K in each $\alpha$ assumption (triangle up for temperature increased, triangle down for decreased). BM$\_$N (COR with new errors) seismic parameters and spectroscopic parameters derived from RAVE spectra and asteroseismology.  }
              \label{Fig:Benoitalltests}
    \end{figure*}

\section{PARAM tensions and additional results}
\label{app:tensions}
S1 is an exemplary case of how an erroneus temperature determination leads to misleading age and mass values using PARAM. In the case of RAVE spectra the spectroscopically determined \Teff is 300 K higher than the temperature derived from the high-resolution spectrum. However, as visible in Fig.~\ref{Fig:tensions}, the erroneus \Teff lead to tensions between the a-posteriori and the input values of \Dnu, \Numax and \Teff, and to asymmetric PDFs. This does not happen when adopting the atmospheric parameters coming from high-resolution spectroscopy.

   \begin{figure*}
   \centering
   \includegraphics[width=0.98\columnwidth]{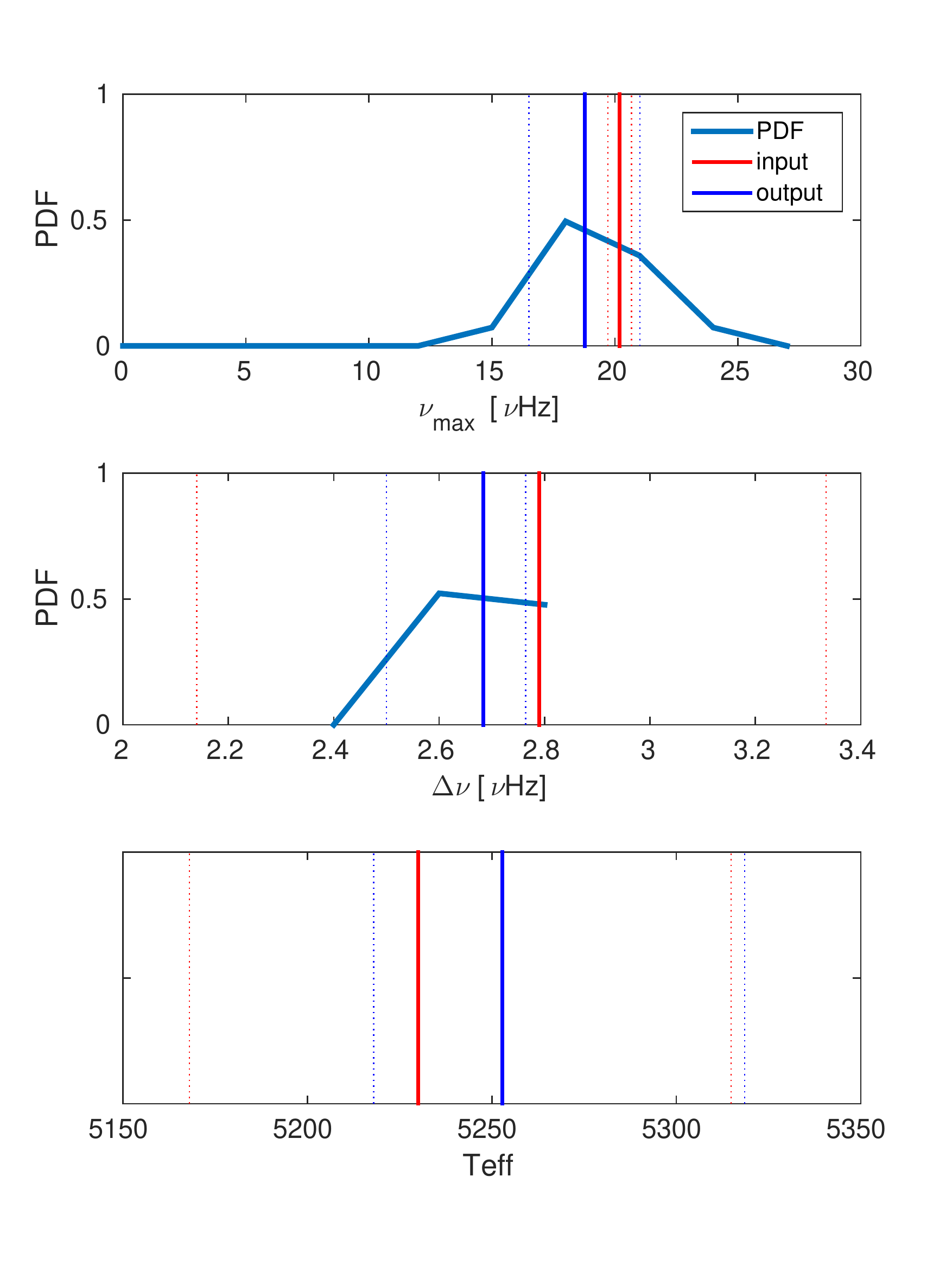}
   \includegraphics[width=0.98\columnwidth]{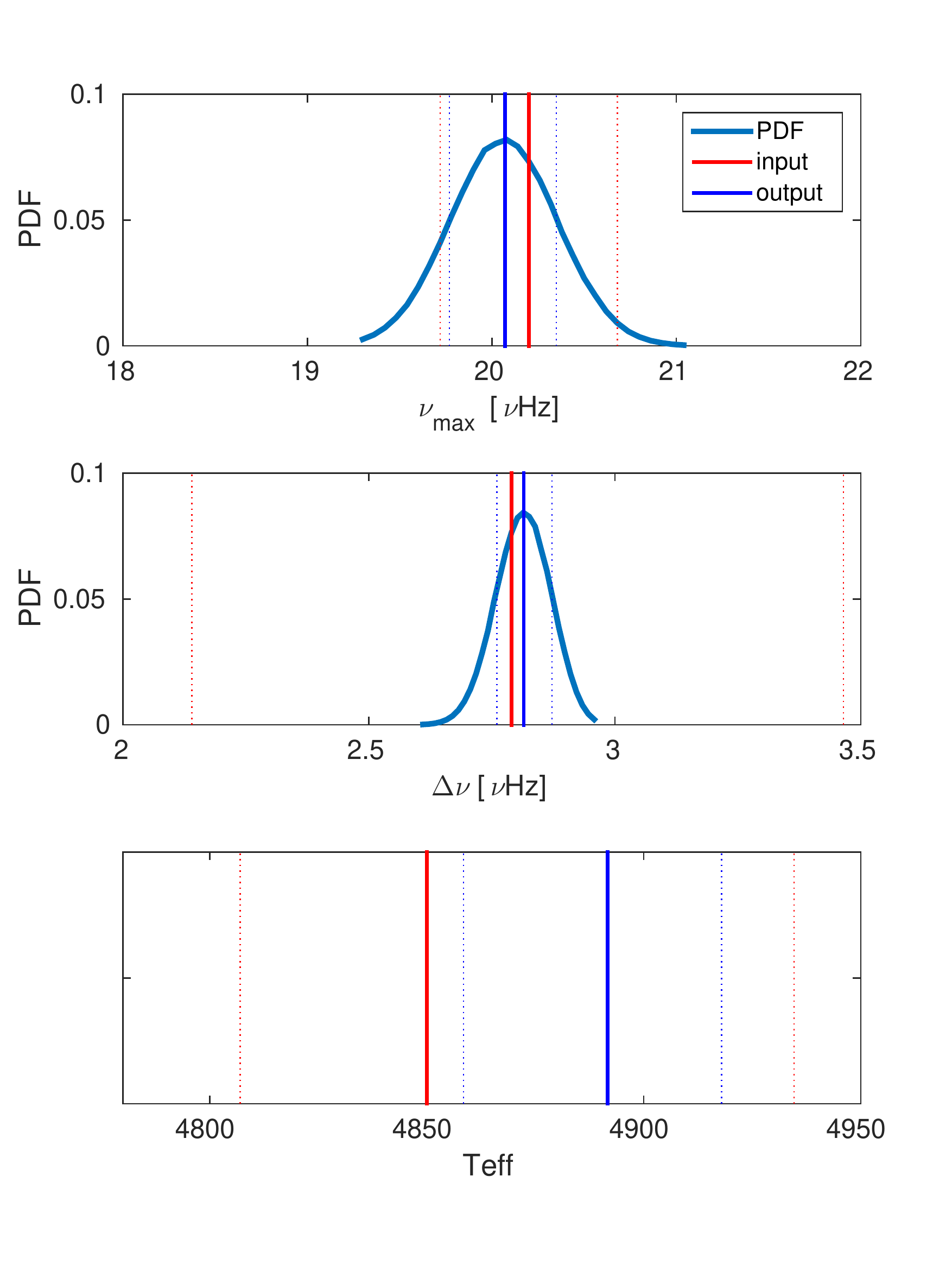}
   \caption{Left column: a posteriori \Dnu, \Numax, and \Teff of S1 using RAVE atmospheric parameters. PDF for \Dnu and \Numax are showed as well. Right column: a posteriori \Dnu (and its PDF), \Numax (and its PDF), and \Teff of S1 using atmospheric parameters derived from UVES spectrum. }
              \label{Fig:tensions}
    \end{figure*}
    
  \begin{figure*}
   \centering
   \includegraphics[width=1.8\columnwidth]{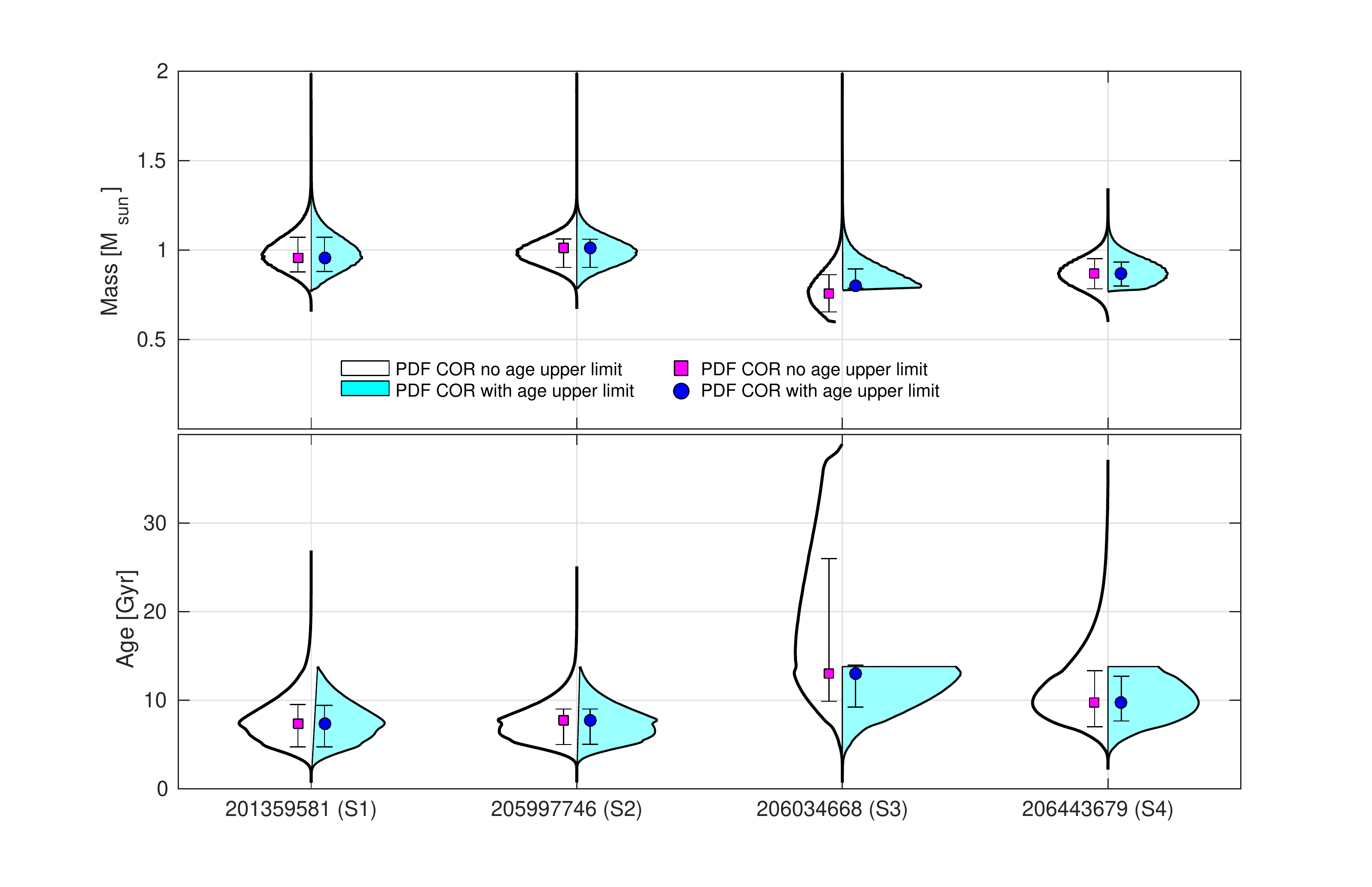}
   \caption{Violin plot of the PDFs of mass (top) and age (bottom) with input seismic parameters given by the COR pipeline with and without upper age limit (cyan and white shade respectively). The mode of each pdf, with the errorbar representing the lower and upper 68th percentile of the PDF, is also indicated.}
   \label{Fig:compAgeLim}
    \end{figure*}

\begin{table*}
\centering
\begingroup
\renewcommand{\arraystretch}{1.2} 
\caption{Ages and masses of the four RAVE stars as derived by PARAM, using COR seismic pipeline and stellar parameters obtained from different spectra, RAVE and ESO-UVES (after adopting the strategy of using seismic gravities to find a more self-consistent surface temperature). For the PARAM results obtained from RAVE spectra, we listed the values corresponding to an $\alpha$ enhancement closer to the measured one. Maximum and minimum error values of age and mass (measured on the 68 percentile of the PDF) are listed in superscript and subscript respectively. COR$_2\sigma{\rm Teff}$ and COR$_{\rm age lim}$ rows list the mass and age determined by doubling the error on \Teff and adding the upper limit on age (13.96 Gyr) respectively.}
\label{Tab:PARAMpipelines}
\begin{tabular}{llcc|cc|cc}
\hline \hline         
Star& Seismic                & \multicolumn{2}{c|}{RAVE}                     & \multicolumn{2}{|c|}{UVES}                      & \multicolumn{2}{|c}{UVES+GAIA}                \\ 
ID  & Pipeline               &Age                     & Mass                  & Age                    & Mass                   & Age                    & Mass                 \\
    &                        &[Gyr]                   & [\Msun]               & [Gyr]                  & [\Msun]                & [Gyr]                  & [\Msun]               \\ \hline 
S1  & COR                    &1.77$^{+0.44} _{-0.46}$ &1.46$^{+0.11} _{-0.15}$&7.42$^{+2.12}_{-2.68}$  &0.96$^{+0.11} _{-0.08}$ & 7.22$^{+1.42}_{-2.60}$ & 1.00$^{+0.08}_{-0.09}$   \\
    & COR$_2\sigma{\rm Teff}$&                        &                       &7.07$^{+1.57}_{-2.97}$  &1.00$^{+0.12} _{-0.09}$ &                        &                       \\
    & COR$_{\rm age lim.}$   &1.91$^{+2.79} _{-0.34}$ &1.13$^{+0.35} _{-0.00}$&7.41$^{+2.03}_{-2.68}$  &0.96$^{+0.11} _{-0.08}$ &                        &                        \\
    & BM\_N                  &                        &                       &25.58$^{+7.99}_{-15.31}$&0.64$^{+0.13} _{-0.04}$ &                        &                        \\ \hline

S2  & COR                    &7.64$^{+1.80} _{-2.58}$ &0.97$^{+0.09} _{-0.07}$&7.76$^{+1.24}_{-2.74}$  &0.99$^{+0.08} _{-0.07}$ & 7.79$^{+1.56}_{-2.36}$ & 0.96$^{+0.08}_{-0.07}$   \\
    & COR$_2\sigma{\rm Teff}$&                        &                       &7.80$^{+2.08}_{-2.47}$  &0.94$^{+1.06} _{-0.06}$ &                        &                        \\
    & COR$_{\rm age lim.}$   &7.61$^{+2.05} _{-2.68}$ &0.99$^{+0.07} _{-0.01}$&7.76$^{+1.25} _{-2.73}$ &1.01$^{+0.05} _{-0.11}$ &                        &                         \\
    & BM\_N                  &                        &                       &7.84$^{+6.12}_{-5.50}$  &1.04$^{+0.28} _{-0.26}$ &                        &                         \\ \hline

S3  & COR                    &12.95$^{+11.00} _{-4.02}$&0.80$^{+0.10} _{-0.01}$&13.01$^{+12.99}_{-3.15}$&0.78$^{+0.11} _{-0.10}$ & 8.18$^{+3.93}_{-2.36}$ & 0.94$^{+0.09}_{-0.11}$   \\
    & COR$_2\sigma{\rm Teff}$&     --                 &    --                 &13.13$^{+0.83} _{-4.17}$&0.79$^{+0.11} _{-0.01}$ &                        &                         \\
    & COR$_{\rm age lim.}$   &12.95$^{+0.89} _{-4.28}$&0.81$^{+0.15} _{-0.03}$&13.04$^{+0.92} _{-3.82}$&0.80$^{+0.10} _{-0.07}$ &                        &                          \\
    & BM\_N                  &                        &                       &15.98$^{+9.15}_{-5.39}$ &0.75$^{+0.10} _{-0.09}$ &                        &                          \\ \hline

S4  & COR                    &10.03$^{+2.70} _{-2.84}$&0.85$^{+0.08} _{-0.06}$&9.58$^{+3.68} _{-2.57}$ &0.87$^{+0.06} _{-0.07}$ & 8.88$^{+2.91}_{-1.88}$ & 0.88$^{+0.08}_{-0.06}$    \\
    & COR$_2\sigma{\rm Teff}$&     --                 &   --                  &9.04$^{+3.69} _{-2.28}$ &0.87$^{+0.10} _{-0.07}$ &                        &                          \\
    & COR$_{\rm age lim.}$   &8.31$^{+3.58} _{-2.85}$ &0.92$^{+0.09} _{-0.11}$&9.72$^{+3.00} _{-2.05}$ &0.87$^{+0.06} _{-0.07}$ &                        &                          \\
    & BM\_N                  &                        &                       &9.72$^{+3.91}_{-2.69}$  &0.87$^{+0.08} _{-0.09}$ &                        &                          \\ 

\hline                                        
\end{tabular}
\endgroup
\end{table*}

\section{M4 PARAM results in detail}
The globular cluster M4 is the ideal testing ground for investigating the accuracy of our stellar age and mass determination respect to other classic techniques. The cluster had been well investigated in literature, and its age has been determined using both CMD fitting \citep[e.g.,][]{Miglio2016} and the white dwarfs cooling sequence \citep{Hansen2004}: both techniques agree on an age of $\sim$13 Gyr with an error of 0.7 Gyr. The work of \citet{Miglio2016} determined also a typical mass of the stars in the RGB: M$_{\rm RGB}$=0.84 \msun, with an error of 0.05 \msun.

In Fig.\ref{Fig:M4pdfs} we report the individual mass and ages PDFs as determined using PARAM and we compare them with the literature results for M4. Our values are in a very good agreement with literature values, with the exception of star M4-S4, a probable red clump star.

  \begin{figure*}
   \centering
   \includegraphics[width=1\columnwidth]{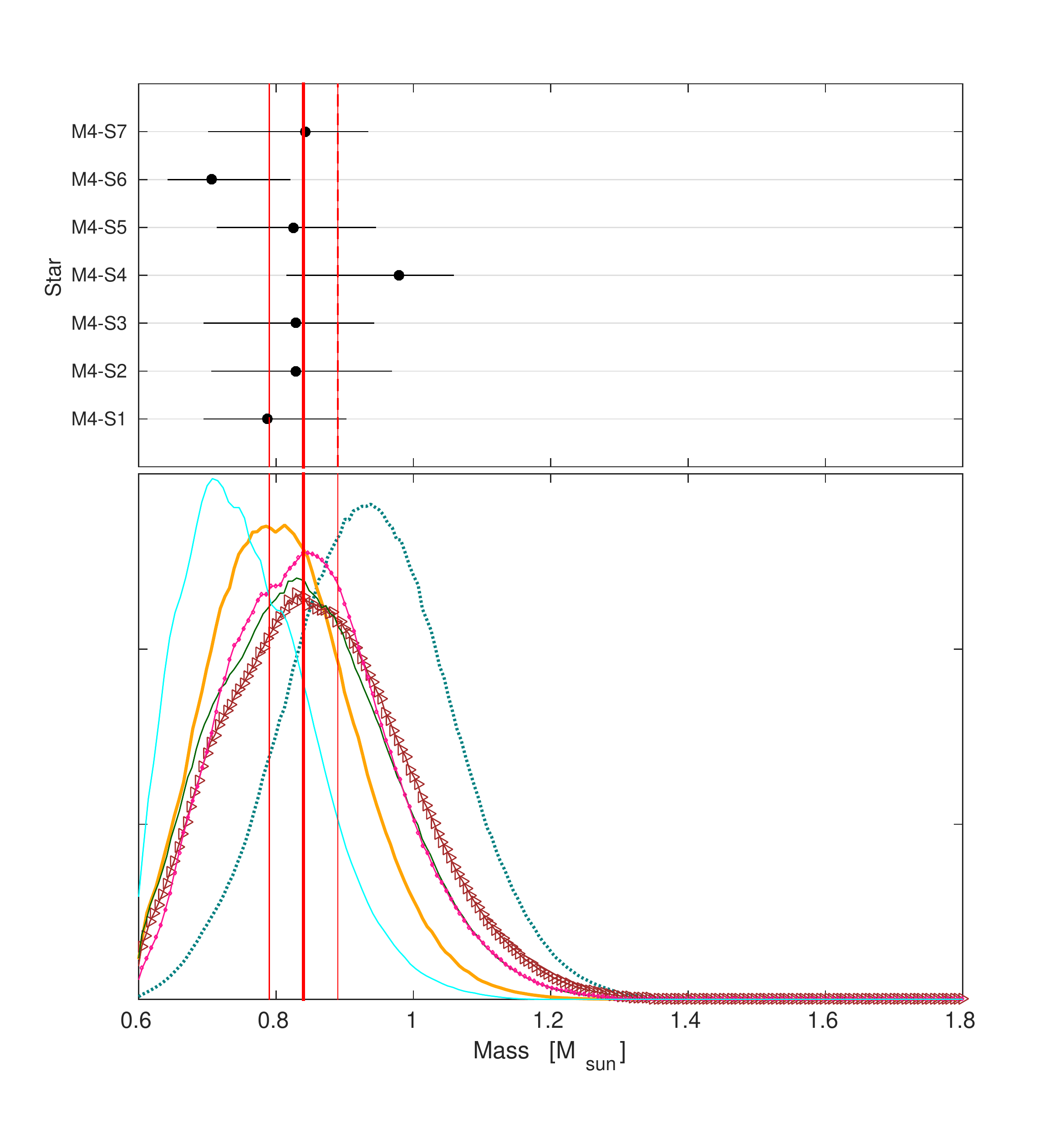}
   \includegraphics[width=1\columnwidth]{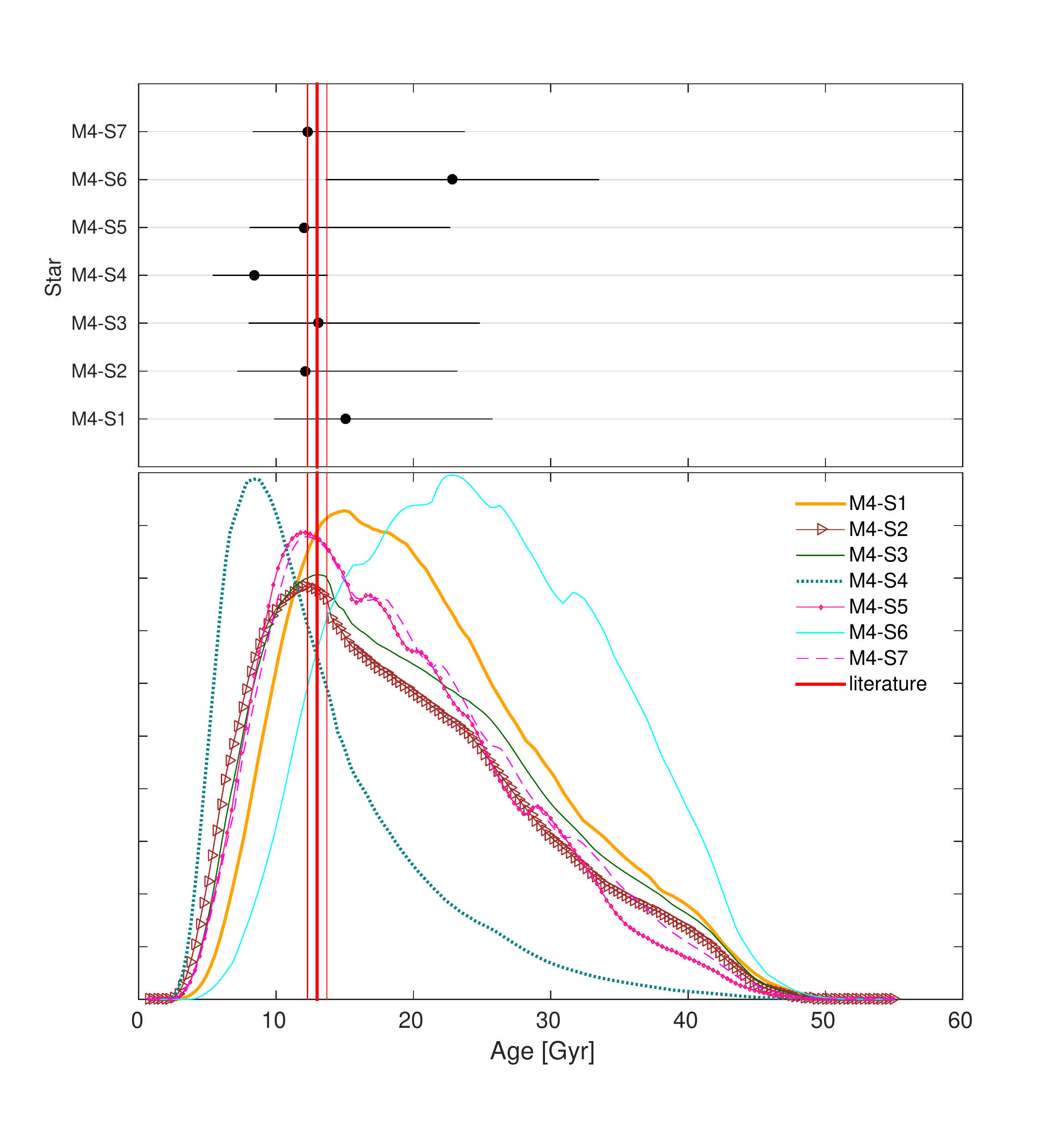}   
   \caption{Masses (left column) and ages (right column) of the red giants in M4 analysed in this work. The thick red vertical line identifies the literature value, while the fine lines identifies the upper an lower values. On the top row are reported individual masses (left) and ages (right), with the errorbar indicating the 68 percentile of the PDF. At the bottom the individual PDF for mass and ages are plotted. }
   \label{Fig:M4pdfs}
    \end{figure*}

\end{appendix}
\end{document}